\documentclass[showpacs,amssymb,10pt,reprint,aps,prd,longbibliography,nofootinbib,floatfix,superscriptaddress]{revtex4-1}
\usepackage{graphicx,epsfig,amssymb} 
\usepackage{amsmath,amsfonts, times}
\usepackage{bm} 
\usepackage[normalem]{ulem}
\usepackage{epstopdf}
\usepackage[caption=false]{subfig}
\usepackage[usenames]{color} 
\usepackage{mathrsfs} 
\usepackage{natbib}
\usepackage{soul}
\usepackage{subfig}
\usepackage[utf8x]{inputenc}
\usepackage{tikz}
\usetikzlibrary{decorations.pathmorphing}
\definecolor{coolblack}{rgb}{0.0, 0.18, 0.39}
\definecolor{darkred}{rgb}{0.5,0,0}
\definecolor{darkgreen}{rgb}{0,0.5,0}
\definecolor{darkblue}{rgb}{0,0,0.5}
\definecolor{lapislazuli}{rgb}{0.15, 0.38, 0.61}
\definecolor{venetianred}{rgb}{0.78, 0.03, 0.08}
\definecolor{bleudefrance}{rgb}{0.19, 0.55, 0.91}
\definecolor{dogwoodrose}{rgb}{0.84, 0.09, 0.41}

\newcommand\numberthis{\addtocounter{equation}{1}\tag{\theequation}}
\newcommand{\xst}{x_{\star}}

\begin{document}
	\title{\large Asymmetric wormholes in Palatini $f(\mathcal{R})$ gravity: Energy conditions, absorption and quasibound states}
	
	\author{Renan B. Magalh\~aes}
	\email{renan.batalha@ext.uv.es}
	\affiliation{Programa de P\'os-Gradua\c{c}\~{a}o em F\'{\i}sica, Universidade 
		Federal do Par\'a, 66075-110, Bel\'em, Par\'a, Brazil.}
	\affiliation{Departamento de F{\'i}sica Te{\'o}rica and \textit{IFIC}, Centro Mixto Universitat de Val\`encia - \textit{CSIC}. Universitat de Val\`encia, Burjassot-46100, Val\`encia, Spain.}	
	\author{Andreu Mas{\'o}-Ferrando}
	\email{andreu.maso@uv.es}
	\affiliation{Departamento de F{\'i}sica Te{\'o}rica and \textit{IFIC}, Centro Mixto Universitat de Val\`encia - \textit{CSIC}. Universitat de Val\`encia, Burjassot-46100, Val\`encia, Spain.}	
	\author{Gonzalo J. Olmo}
	\email{gonzalo.olmo@uv.es}
	\affiliation{Departamento de F{\'i}sica Te{\'o}rica and \textit{IFIC}, Centro Mixto Universitat de Val\`encia - \textit{CSIC}. Universitat de Val\`encia, Burjassot-46100, Val\`encia, Spain.}
	\affiliation{Universidade Federal do Cear\'a (UFC), Departamento de F\'isica,\\ Campus do Pici, Fortaleza - CE, C.P. 6030, 60455-760 - Brazil.}
	\author{Lu\'is C. B. Crispino}
	\email{crispino@ufpa.br}
\affiliation{Programa de P\'os-Gradua\c{c}\~{a}o em F\'{\i}sica, Universidade Federal do Par\'a, 66075-110, Bel\'em, Par\'a, Brazil.}%
\affiliation{Departamento de Matem\'atica da Universidade de Aveiro and Centre for Research and Development  in Mathematics and Applications (CIDMA), Campus de Santiago, 3810-183 Aveiro, Portugal.}
	
	\begin{abstract}
We investigate the scalar absorption spectrum of wormhole solutions constructed via the recently developed thin-shell formalism for Palatini $f(\cal R)$ gravity. Such wormholes come from the matching of two Reissner-Nordström spacetimes at a time-like hypersurface (shell), which, according to the junction conditions in Palatini $f(\cal R)$, can be stable and have either positive or negative energy density. In particular, we identified a new physically interesting configuration made out of two overcharged Reissner-Nordström spacetimes, whose absorption profile departs from that of black holes and other previously considered wormholes in the whole range of frequencies. Unlike in symmetric wormhole solutions, the asymmetry of the effective potential causes the dilution of the resonances associated to the quasibound states for the high-frequency regime. Therefore, slight asymmetries in wormhole space-times could have a dramatic impact on the observable features associated to resonant states.
\end{abstract}
	\date{\today}
	\maketitle

	\section{Introduction}\label{sec:int}
Though wormholes are generally regarded as exotic geometric objects, they are gaining increasing attention in theoretical physics~\cite{maldacena:2013}, with efforts made to understand and simulate their features~\cite{visser:1995,krasnikov:2008,jusufi:2018, jafferis:2022},  and also to characterise their observational  signatures~\cite{Dai:2019mse,De Falco:2020,bambi:2021,Simonetti:2020ivl}. The chronicles of wormholes began to be written in 1935 with the seminal paper of Einstein and Rosen~\cite{einstein:1935}, where they were used as a geometrical model  that could avoid some undesired features of point particles.

The revival of wormholes in physics came with the conception of traversable wormholes in General Relativity~\cite{ellis:1973,morris:1988a,morris:1988,visser:1989}, which could provide a way to travel to distant places (and times). The price to pay for traversability was the violation of some energy conditions~\cite{visser:1995,visser:2003}, which requires exotic matter sources. 
In order to construct that new class of wormholes and minimise the amount of exotic matter, one can apply the thin-shell formalism~\cite{israel:1966}, grafting two spacetimes at a hypersurface, giving rise to a geodesically complete space-time where the energy conditions are violated only in a small region \cite{visser:1989b,poisson:1995,ishak:2002, eiroa:2004}.

More recently, and motivated by an astrophysical interest, different kinds of wormholes and other ultra compact objects have been studied as black hole mimickers~\cite{mielke:2000,visser:2004,damour:2007,bambi:2016,Konoplya:2016,Battista:2017,rosa:2018,Afonso:2019fzv,De Falco:2021,rosa:2021,Konoplya:2022}, since they can share similarities with Schwarzschild/Kerr-like objects~\cite{lemos:2008,herdeiro:2021}. The reason is that these objects can present features that allow to tell them apart from their black hole cousins, and future measurements of gravitational waves, and advances in very-long-baseline interferometry could, in principle, find some characteristic imprints of them~\cite{cunha:2015,aneesh:2018,johnson-mcdaniel:2020}. The case of wormholes is particularly interesting, because the possibility of having large amounts of exotic energy sources at their throats offers a unique opportunity to study new phenomenology that could affect aspects such as \textit{shadows}~\cite{wielgus:2020,wang:2020,GORG:2021}, gravitational waves ringdown \textit{echoes}~\cite{cardoso:2016,cardoso:2016b,konoplya:2019} and, in general, the propagation of waves and quantum fields in those regions. We will focus in the latter aspect, considering the propagation of scalar waves in a thin-shell wormhole background.

Absorption and scattering of particles and fields by black holes~\cite{matzer:1968,fabbri:1975,unruh:1976,CHM:2010,OCH:2011,CDHO:2014,CDHO:2015,LDC:2017,LDC:2018,BC:2019,paula:2020,MLC:2020EPJC,MLC:2022EPJC} 
and ultra compact objects~\cite{macedo:2018} has been studied in the literature over the years also in an effort to improve our understanding of the spectroscopy of compact objects ~\cite{cabero:2020}. In particular, fields living in the vicinity of compact objects may give rise to phenomena such as the emergence of quasibound states~\cite{macedo:2018,chandrasekhar:1991,cardoso:2014}, clouds~\cite{benone:2014,benone:2015}, and superradiant scattering~\cite{BC:2016,BC:2019}. Wormholes are among the objects that might have quasibound states around them. As a consequence, this may create resonances in their absorption spectra~\cite{delhom:2019,limajr:2020}, and also change their ringdown profile~\cite{cardoso:2016,Churilova:2021}. Exploring how these properties are modified in configurations characterized by positive and negative energy densities is the main goal of this paper. In addition, this allows to investigate whether the absorption spectrum is sensitive to the sign of the energy density at the throat. A massless scalar field represents the simplest quantum probe that one may consider and, at the same time, it offers basic phenomenology that one may expect to occur in more complex field distributions.

Besides their existence, the stability of solutions is a fundamental aspect to have into account when modeling exotic objects~\cite{visser:1989b,poisson:1995}. In this regard, stable thin-shell wormholes in four dimensional General Relativity generically require negative energy densities. Though this may not be a problem in an accelerating expanding universe thought to be driven by some kind of exotic energy source with repulsive gravitational properties~\cite{Frieman:2008,Caldwell:2009,Weinberg:2013}, it is always desirable to find stable solutions that do not necessarily require negative energies. It has been recently shown that this is indeed the case in $f(\cal R)$ extensions in the Palatini formulation~\cite{ORG:2020,LOORGR:2020}~, where stable wormhole solutions can be generically found in thin-shell scenarios with positive and negative energy densities. Moreover, due to the peculiarities of their thin-shell equations, that property is independent of the particular choice of the $f(\cal R)$ function, which contrasts with the purely metric formulation of $f(R)$ theories. In the metric $f(R)$ case, the thin-shell dynamics depends on the specific $f(R)$ function chosen~\cite{Senovilla:2013}, thus making any analysis strongly model dependent. For this reason, in this paper we focus on thin-shell wormholes constructed in the Palatini formulation of $f(\cal R)$. Nonetheless, it is important to point out that the cut-and-paste procedure can also produce thin-shell wormholes supported by positive energy matter in other modified theories of gravity~\cite{gravanis:2007,harko:2013,moraes:2018}, and in General Relativity in higher dimensions~\cite{svitek:2018}.

In the Palatini formalism, connection and metric are regarded as independent geometrical objects~\cite{Ferraris:1982}. In the General Relativity case, this has no impact at all in the field equations, but for $f(\cal R)$ extensions the difference is certainly relevant~\cite{olmo:2005a,olmo:2005b}. Unlike other theories of modified gravity, the Palatini dynamics exhibits nonlinear contributions induced by the matter sources, having no new dynamical degrees of freedom. This allows them to rather generically satisfy current solar system constraints and also be compatible with recent gravitational wave astronomy results~\cite{Olmo:2011uz,Olmo:2019flu}, though some exceptions also exist. Furthermore, its versatility and effectiveness have been proven in a wide range of scenarios and scales \cite{Olmo:2019flu,Wojnar:2022txk,Maso-Ferrando:2021ngp}.

We study the propagation of a massless scalar field in a background constructed by gluing together two Reissner-Nordström solutions. The junction conditions for these configurations correspond to those of Palatini $f(\cal R)$ theories, which involve a thin-shell that stabilizes the solution for a certain range of positive or negative energy densities, depending on model parameters. As a result, we can deal with symmetric and asymmetric wormholes that represent electrovacuum spacetimes, which can have the same or different charge and mass on each side. The asymmetric configurations will be referred to as RN-AWH, which stands for Reissner-Nordström asymmetric wormhole.

The content of this paper is organized as follows. In Sec.~\ref{sec:AWH} we review the construction of the RN-AWH in Palatini $f(\cal R)$ gravity, and specify the parameter space that we will consider in the following sections. The absorption problem is described in Sec.~\ref{sec:abs}, where we present a selection of our numerical results and provide a discussion about the emergence of resonant peaks in the absorption cross section.  Finally, we summarize our results
and discuss some perspectives in Sec.~\ref{sec:con}.

	\section{Asymmetric wormholes in Palatini $f(\cal R)$}\label{sec:AWH}
	\subsection{Framework: spacetime surgery}
	The spacetime surgery is a well-known technique to construct wormhole spacetimes~\cite{visser:1989,visser:1989b} -- which are called thin-shell wormholes. In 4-dimensions General Relativity, these objects are constructed in such a way that they violate the energy conditions just in a thin layer of the manifold. However, modified theories of gravity and also General Relativity in higher dimensions enable the construction of thin-shell wormholes supported by non-exotic matter~\cite{LOORGR:2020,gravanis:2007,harko:2013,moraes:2018,svitek:2018}.  The construction follows from the \textit{cut and paste} procedure. By considering two smooth manifolds, say $\mathcal{M}_{\pm}$ (with associated metrics $g^{\pm}_{\mu\nu}$), one may \textit{cut} them so that each one becomes bounded by a time-like surface $\Sigma_{\pm}$. After that, one may \textit{paste} them together at their boundary time-like surfaces, producing a single manifold $\mathcal{M}=\mathcal{M}_{-}\cup\mathcal{M}_{+}$ with a thin hypersurface $\Sigma=\Sigma_{\pm}=\mathcal{M}_{-}\cap\mathcal{M}_{+}$ that connects two regions with geometries governed by $g_{\mu\nu}^{-}$ and $g_{\mu\nu}^{+}$. In fact, across the hypersurface $\Sigma$ several discontinuities on geometric and matter quantities may exist~\cite{clarke:1987}, then one needs to use a suitable  framework to describe these structures (one may use tensorial distributions instead of tensorial functions). In essence, the geometric and matter quantities must satisfy at the hypersurface $\Sigma$ the so-called \textit{junction conditions}~\cite{mars:1993}. Among them, the simplest one is to require the metric to be continuous across the hypersurface, $g_{\mu\nu}^{+}|_{\Sigma}=g_{\mu\nu}^{-}|_{\Sigma}$, while the other junction conditions are usually deeply dependent on the chosen gravity model.
	
	Here we will consider the family of $f(\mathcal{R})$ gravity theories constructed \textit{à la} Palatini, i.e, with the spacetime metric and the affine connection being independent gravity fields. Here $f(\mathcal{R})$ denotes a function of the Ricci scalar $\mathcal{R}\equiv g^{\mu\nu}\mathcal{R}_{(\mu\nu)}(\Gamma)$, with $g^{\mu\nu}$ being the contravariant components of the spacetime metric and $\mathcal{R}_{(\mu\nu)}(\Gamma)$ the symmetric part of the Ricci tensor, constructed only as a function of the affine connection $\Gamma$ (with components $\Gamma^{\alpha}_{\mu\nu}$). The action of the $f(\mathcal{R})$ model is
	\begin{equation}
		\label{eq:grav_model}
		\mathcal{S} = \dfrac{1}{2\kappa^2}\int d^4 x \sqrt{-g}f(\mathcal{R}) \,+\int d^4 x \sqrt{-g} \mathcal{L}_{m}(g_{\mu\nu},\psi_m),
	\end{equation}
	where $\kappa^2$ is the gravitational constant in suitable units, $g$ is the metric determinant and $\mathcal{L}_{m}$ is the matter Lagrangian, which depends on the metric and matter fields. In the Palatini picture, by varying the action~\eqref{eq:grav_model} with respect to the metric and the affine connection, one finds the field equations
	\begin{align}
		\label{eq:fieldeq_metric} f_\mathcal{R} \mathcal{R}_{(\mu\nu)} - \frac{1}{2} f(\mathcal{R}) g_{\mu\nu} &= \kappa^2 T_{\mu\nu},\\
		\label{eq:fieldeq_connec} \nabla^{\Gamma}_{\lambda}(\sqrt{-g}f_{\mathcal{R}}g^{\mu\nu}) &= 0,
	\end{align}
where $T_{\mu\nu}\equiv \frac{2}{\sqrt{-g}}\frac{\delta (\sqrt{-g}\mathcal{L}_m)}{\delta g^{\mu\nu}}$ is the so-called energy-momentum tensor and $f_\mathcal{R}\equiv df/d\mathcal{R}$. The right-hand side of Eq.~\eqref{eq:fieldeq_connec} is zero because we assume that the matter Lagrangian is independent of the affine connection \cite{Afonso:2017bxr}. Taking the trace of Eq.~\eqref{eq:fieldeq_metric}, one finds that there is an algebraic relation between the Ricci scalar and the matter fields, such that $\mathcal{R}\equiv\mathcal{R}(T)$, where $T$ is the trace of the energy-momentum tensor. This relation, together with the bi-metric structure introduced by Eq.~\eqref{eq:fieldeq_connec} ($\Gamma$ is Levi-Civita of an auxiliary metric $q_{\mu\nu} \equiv f_{\mathcal{R}}(T)g_{\mu\nu}$), allows one to interpret the Palatini $f({\mathcal{R}})$ as similar to General Relativity plus additional couplings in the matter fields.

	In order to analyse the \textit{glued} spacetimes, we move to a consistent mathematical framework to study geometric and matter fields, i.e, we start to consider tensorial distributions instead of tensorial functions. In this approach, the metric and energy-momentum distributions can be written as~\cite{clarke:1987}
	
	\begin{align}
		\label{eq:metric_distr} \underline{g}{_{\mu\nu}} &= g^{+}_{\mu\nu}\underline{\theta}+g^{-}_{\mu\nu}(\underline{1}-\underline{\theta}),\\
		\label{eq:em_distr} \underline{T}{_{\mu\nu}} &= T^{+}_{\mu\nu}\underline{\theta}+T^{-}_{\mu\nu}(\underline{1}-\underline{\theta})+S_{\mu\nu}\underline{\delta}^{\Sigma},
	\end{align}
	where underlined quantities denote distributions. In Eq.~\eqref{eq:metric_distr} we used the continuity condition of the metric across the hypersurface. $T^{\pm}_{\mu\nu}$ are the energy-momentum tensors on each side of the hypersurface, respectively $\mathcal{M}^{\pm}$; $\underline{\theta}$ is the Heaviside step function, which takes the value 1 in  $\mathcal{M}^{+}$, 0 in $\mathcal{M}^{-}$ and any reference value on the hypersurface; $\underline{\delta}^{\Sigma}$ is a Dirac's delta-type distribution with support on the hypersurface, defined by $<\underline{\delta}^{\Sigma},X>\equiv \int_{\Sigma}X$, for any function $X$; and $S_{\mu\nu}$ is the singular part of the energy-momentum tensor on the hypersurface. Analogously to Eq.~\eqref{eq:em_distr}, the distributional form of the trace of the energy-momentum tensor is given by
	\begin{equation}
		\label{eq:tr_distr}
		\underline{T} = T^+\underline{\theta}+T^{-}(\underline{1}-\underline{\theta})+S\underline{\delta}^{\Sigma},
	\end{equation}
	where $S = S^{\rho}_{\rho}$ is the trace of the singular part of the energy-momentum tensor.
	
	As mentioned above, although the metric is continuous (but not differentiable) across the hypersurface, other curvature and matter distributions are not. In order to identify the \textit{allowed} discontinuities of these quantities across the hypersurface, one has to make use of the junction conditions, which introduces constrains in the discontinuities of curvature and matter quantities in both sides of the hypersurface. These junction conditions are highly influenced by the considered gravitational framework, and in alternative theories of gravity these conditions may change significantly from General Relativity. Considering the Palatini $f(\mathcal{R})$ framework, the junction conditions are
	\begin{align}
		\label{eq:jc_1}&[g_{\mu\nu}]=0 \text{ and } [h_{\mu\nu}]=0,\\
		\label{eq:jc_2}&[T]=0 \text{ and } S=0,\\
		\label{eq:jc_3}&\dfrac{1}{3}h_{\mu\nu}[K^{\rho}_{\rho}] -[K_{\mu\nu}]= \kappa^2 \dfrac{S_{\mu\nu}}{f_{\mathcal{R}|_\Sigma}},\\
		\label{eq:jc_4}&D^{\rho}S_{\rho \nu}= -n^{\rho}h^{\sigma}_{\nu}[T_{\rho\sigma}],\\
		\label{eq:jc_5}&\left(K^{+}_{\rho\sigma}+K^{-}_{\rho\sigma}\right)S^{\rho\sigma} = 2n^{\rho}n^{\sigma}[T_{\rho\sigma}] - \dfrac{3\mathcal{R}^2_{T}f_{\mathcal{RR}}^2}{f_{\mathcal{R}}}[b^2],
	\end{align}
where the brackets denote discontinuity of the quantity inside them, across the hypersurface $\Sigma$, i.e, $[A] \equiv A^{+}|_{\Sigma}-A^{-}|_{\Sigma}$. (For details in the derivation of these junction conditions see Ref.~\cite{ORG:2020}.)
$h_{\mu\nu} = g_{\mu\nu}-n_{\mu}n_{\nu}$ is the pullback of the first fundamental form (the induced metric on the hypersurface $\Sigma$), with $n^{\mu}$ being the unit vector normal to $\Sigma$, and $K_{\mu\nu}^{\pm}\equiv h^{\rho}_{\mu}h^{\sigma}_{\nu}\nabla^{\pm}_{\rho}n_{\sigma}$ is the pullback of the second fundamental form (the extrinsic curvature). In the last two junction conditions, $D^{\rho}\equiv h_{\alpha}^{\rho}\nabla^{a}$ is the covariant derivative on the hypersurface, $\mathcal{R}_{T}\equiv d\mathcal{R}/dT$ and $b\equiv n^{\mu}[\nabla_{\mu} T]$.

In General Relativity, both the metric and Palatini formalisms result in the same set of field equations. However, by considering a $f(R)$ Lagrangian, the metric and Palatini approaches lead to completely different sets of equations of motion. Consequently, the junction conditions of Palatini $f(\cal R)$ largely depart from the corresponding expressions in General Relativity and in the metric version of $f(R)$. A remarkable aspect of the Palatini $f(\cal R)$ junction conditions is the vanishing of brane tension, $S$. In the framework of General Relativity, one has $h_{\mu\nu}[K^{\rho}_{\rho}] -[K_{\mu\nu}]= \kappa^2 S_{\mu\nu}$ instead of Eq.~\eqref{eq:jc_3}, and the brane tension in general is non-vanishing, $\kappa^2 S = 2[K^{\rho}_{\rho}]$. In the framework of Palatini $f(\cal R)$ it does happen regardless of the behavior of $[K^{\rho}_{\rho}]$.
	
	\subsection{Asymmetric RN-RN wormholes}
	With the junction conditions~[Eqs. (\ref{eq:jc_1})-(\ref{eq:jc_5})] one can match two static and spherically symmetric spacetimes $\mathcal{M}_{\pm}$ on a given hypersurface $\Sigma$, constructing a wormhole that connects two regions by a throat. The line elements of each side of the throat can be written as
	\begin{equation}
		\label{eq:line_elements} ds^2_{\pm} = -A_\pm(r_\pm)dt^2 + B^{-1}_\pm(r_\pm)dr_\pm^2+r_\pm^2d\Omega^2,
	\end{equation}
	where $d\Omega = d\theta^2 + \sin^2\theta d\phi^2$ is the line element of a unit sphere and $A_\pm(r_\pm)$ and $B_\pm(r_\pm)$ are the metric functions on each side of the throat, which depend only on $r_\pm$, i.e,  the radial coordinate on each side. The hypersurface $\Sigma$ has coordinates $x^{\mu} = (t,R,\theta,\phi)$, where $r_\pm=R$ is the areal radius of it. One can parametrize this hypersurface in terms of the proper time $\tau$ of an observer comoving to it. Hence, the line element on $\Sigma$ can be written as~\cite{ishak:2002}
	\begin{equation}
		\label{eq:metric_hypersurface}
		ds^2_\Sigma = -d\tau^2+R^2(\tau)d\Omega^2.
	\end{equation}
	The tangent vectors to the hypersurface are $e^{\mu}_{\theta} = (0,0,1,0)$, $e^{\mu}_{\phi} = (0,0,0,1)$ and $U^{\mu} = (\dot{t},\dot{R},0,0)$, while the unit vector normal to $\Sigma$ on each side is $n^{\mu\pm} = \pm(\dot{R}/\sqrt{AB},\sqrt{B+\dot{R}^2},0,0)$, where the overdot denotes derivatives with respect to the proper time, and the functions $A=A_{\pm}(R)$ and $B=B_{\pm}(R)$ are the metric functions evaluated at $r_{\pm}=R$.
	
	One can compute the components of the extrinsic curvature on each side of the hypersurface (in general they are not equal, since although the metric is continuous over $\Sigma$, its derivative is not) via $K_{ij}^\pm = e^\mu_i e^\nu_j\nabla^{\pm}_{\mu}n_\nu$~\cite{ORG:2020}. Therefore, the non-vanishing components of the second fundamental form are $K^{i}{_j}{^\pm} =\text{diag}(K^{\tau}{_\tau}{^\pm},K^{\theta}{_\theta}{^\pm},K^{\theta}{_\theta}{^\pm})$, with~\cite{GORG:2021}
	\begin{align}
		\label{eq:Ktautau} K^{\tau}{_\tau}{^\pm} &= \pm\dfrac{B_{\pm}^2A_{R\pm}+(B_{\pm}A_{R\pm}-A_{\pm}B_{R\pm})\dot{R}^2+2A_{\pm}B_{\pm}\ddot{R}}{2A_{\pm}B_{\pm}\sqrt{B_{\pm}+\dot{R}^2}},\\
		\label{eq:Ktetateta} K^{\theta}{_\theta}{^\pm} &= \pm\dfrac{\sqrt{B_{\pm}+\dot{R}^2}}{R},
	\end{align}
	where $A_R \equiv dA/dR$ and $B_R \equiv dB/dR$.
	
	The matter content of the thin shell (the singular part of the energy-momentum tensor) can be modelled  as a perfect fluid distribution, i.e, $S^\mu{_\nu} = \text{diag}(-\sigma,\mathcal{P},\mathcal{P})$, where $\sigma$ and $\mathcal{P}$ are the surface energy density and the tangential surface pressure density, respectively. Due to Eq.~\eqref{eq:jc_2} one finds that the pressure density $\mathcal{P}=\sigma/2$ is fully determined by the energy density $\sigma$ (particularly inheriting its sign), hence in Palatini $f(\mathcal{R})$ no equation of state $\mathcal{P}=\mathcal{P}(\sigma)$ is required to close the system, which contrasts with General Relativity and metric $f(\mathcal{R})$~\cite{LOORGR:2020}, and the number of effective degrees of freedom is reduced to just one.
	
	With the junction condition~\eqref{eq:jc_3}, one moves the problem to determine the energy density of the system to compute the difference between the discontinuities of the extrinsic curvature components, i.e,
	\begin{equation}
		\label{eq:field_eq_shell}
		[K^{\tau}{_\tau}]-[K^{\theta}{_\theta}] = \dfrac{3\kappa^2}{2f_{\mathcal{R}|_\Sigma}}\sigma.
	\end{equation} 
	Finally, we look for the energy conservation relation~\eqref{eq:jc_4}, which in the spherically symmetric case reduces to
	\begin{equation}
		\label{eq:en_cons}
		-D^{\rho}S_{\rho\nu} = \left[\dot{\sigma}+\dfrac{2\dot{R}}{R}(\sigma+\mathcal{P})\right]\delta^{\tau}{_\nu} = n^\rho h^\sigma{_\nu}[T_{\rho\sigma}],
	\end{equation}
	where $\delta^{\tau}{_\nu} = (1,0,0)$. Using the relation between the pressure and energy densities, one finds that
	\begin{equation}
		\label{eq:en_cons_sig}
		\dfrac{1}{R^3}\dfrac{d\left(\sigma R^3\right)}{d\tau}\delta^{\tau}{_\nu} = n^\rho h^\sigma{_\nu}[T_{\rho\sigma}],
	\end{equation}
	which leads to simple solutions for $\sigma$ in the case where its right-hand side  vanishes, namely $\sigma=C/R^3$, where $C$ is an integration constant. Fortunately, in the electrovacuum scenario this is true. To see it, we recall that for any electrostatic, spherically symmetric field described by a nonlinear electrodynamics, the energy-momentum tensor associated to it can be written as $T_\rho{^\sigma} = \text{diag}(-\phi_1(r),-\phi_1(r),\phi_2(r),\phi_2(r))$, where the functions $\phi_i$ characterize each particular configuration. (In Maxwell electrodynamics, $\phi_i(r)=-q^2/r^4$, with $q$ being the charge per unit mass of the system. In vacuum, $\phi_i=0$.) By contracting the normal vector to the hypersurface with the energy-momentum tensor, one finds that $n^\rho T_{\rho}{^\nu} = -\phi_1(r)n^\nu$, hence the right-hand side of Eq.~\eqref{eq:en_cons_sig} becomes
	\begin{equation}
		\label{eq:hds_en_cos}
		n^\rho h^\sigma{_\nu}[T_{\rho\sigma}]
		=n^\rho h_{\sigma\nu}[T_\rho{^\sigma}] = -(\phi^+_1(r)-\phi^-_1(r))n^{\sigma}h_{\sigma\nu},
	\end{equation} 
	which is identically zero, since
	\begin{equation}
		\label{eq:n_h}
		n^{\sigma}h_{\sigma\nu} = n^{\sigma}\left(g_{\sigma\nu}-n_\sigma n_\nu\right) = n_\nu-n_\nu\equiv 0.
	\end{equation}
	Therefore, any two electrovacuum spacetimes supported by electrostatic and spherically symmetric fields can be glued together at a hypersurface $\Sigma$ with surface energy density $\sigma=C/R^3$. Here, in particular, we are interested in cutting and pasting two Reissner-Nordström (RN) spacetimes, that have different charges and masses, being described by the following line elements
	\begin{equation}
		\label{eq:RN_plus_minus} ds^2_{\pm} = -f_\pm(r_\pm)dt^2 + \dfrac{dr^2_\pm}{f_\pm(r_\pm)}+r^2_\pm d\Omega^2,
	\end{equation}
	with $f_\pm(r_\pm) = 1-2M_\pm/r_\pm+Q^2_\pm/r^2_\pm$, where $M_\pm$ and $Q_\pm$ are the mass and charge on each RN spacetime, respectively. We point out that we do not impose the restriction $Q<M$. For $Q>M$, the line element~\eqref{eq:RN_plus_minus} describes an overcharged RN space-time, which is a naked singularity. We emphasize that all wormholes studied here are geodesically complete, since the matching surface is located beyond where the singularity would be.
	
	One can use the field equation on the shell~\eqref{eq:field_eq_shell} to write~\cite{GORG:2021} 
	
	\begin{equation}
		\label{eq:ddotR}
		\ddot{R} = \dfrac{\gamma - \tfrac{3M_+R-2Q_+^2-R^2(\dot{R}^2+1)}{\sqrt{f_+(R)+\dot{R}^2}}- \tfrac{3M_-R-2Q_-^2-R^2(\dot{R}^2+1)}{\sqrt{f_-(R)+\dot{R}^2}}}{R^3\left(\tfrac{1}{\sqrt{f_+(R)+\dot{R}^2}}+\tfrac{1}{\sqrt{f_-(R)+\dot{R}^2}}\right)},
	\end{equation}
	where $\gamma = 3\tilde{\kappa}^2C/2$ is the energy parameter, with $\tilde{\kappa}^2 = \kappa^2/f_{\mathcal{R}|_\Sigma}$ being a constant, once $\mathcal{R}=\mathcal{R}(T)$ in any Palatini $f(\mathcal{R})$ theory is determined, and we are considering a trace-free energy-momentum tensor. In order to study the linear stability of these (asymmetric) wormhole solutions, one assumes that there is an equilibrium configuration, such that $\dot{R}=0$, and expands Eq.~\eqref{eq:ddotR} in Taylor series around the throat radius of the equilibrium configuration $R_0$~\cite{GORG:2021}, which at first order gives
	\begin{equation}
		\label{eq:exp_ddotR}
		\ddot{R}\approx C_1(R_0)+C_2(R_0)(R-R_0)+\mathcal{O}(R-R_0)^2,
	\end{equation}
	where $C_1$ and $C_2$ are cumbersome functions of $R_0$, $\gamma$, and of the masses and charges of each side. As discussed in Ref.~\cite{GORG:2021}, to have an equilibrium configuration, the first term of the expansion must vanish, and the second one must be negative for a stable equilibrium. 
	
	Before we discuss the stability condition, it will be convenient to introduce a set of dimensionless variables, in order to simplify the expressions, namely
	\begin{align*}
		&r_\pm = x_\pm M_-, && R=x M_-, &&R_0 = x_0 M_-,\\
		& \tau = \tilde{\tau} M_-, &&t = \tilde{t} M_-, &&Q_{-}^2 = y M^2_-,\\
		&Q_{+}^2 = \eta\, Q^2_-, &&M_{+} = \xi M_-, && \gamma =\tilde{\gamma} M_{-}^2,\\
	\end{align*}
where $x_\pm$ are the dimensionless radial coordinates on each side of the throat, $x$ is the dimensionless radius of the throat, $x_0$ is the dimensionless radius of the throat of an equilibrium configuration, $y$ is the charge-to-mass ratio in $\mathcal{M}_-$, $\eta$ gives the relation between the charge content in $\mathcal{M}_+$ and in $\mathcal{M}_-$ (for simplicity, we shall later on refer to $\eta$ as charge-to-charge ratio), $\tilde{\tau}$ and $\tilde{t}$ are dimensionless time variables, and $\xi$ is the mass-to-mass ratio between the two sides, which due to the continuity of the metric across $\Sigma$, must satisfy
	\begin{equation}
		\label{eq:xi}
		\xi = 1 - \dfrac{y}{2x}(1-\eta).
	\end{equation} 
	
	Now we can continue our discussion about the stability of equilibrium solutions. The equilibrium condition ($C_1=0$) leads us to 
	\begin{equation}
		\label{eq:en_par_norm}
		\tilde{\gamma} = -x_0\dfrac{4(x_0-3)x_0+(\eta+7)y}{2\sqrt{(x_0-2)x_0+y}},
	\end{equation} 
	and substituting this expression in Eq.~\eqref{eq:exp_ddotR}, one finds an equation of the form
	\begin{equation}
		\label{eq:stab_eq}
		\dfrac{d^2\delta(\tilde{\tau})}{d\tilde{\tau}^2}+\varpi^2\delta(\tilde{\tau}) = 0,
	\end{equation}
	where $\delta(\tilde{\tau}) \equiv x(\tilde{\tau})-x_0 $ and $\varpi^2$ is given by
	\begin{align*}
		\label{eq:stability}
		\varpi^2 &=-\dfrac{ 4x_0y(\eta-(\eta-7)x_0-17)+((\eta-1)^2+16)y^2}{8x_0^4((x_0-2)x_0+y)}\\
		&-\dfrac{8(2x_0^2-8x_0+9)x_0^2}{8x_0^4((x_0-2)x_0+y)}. \numberthis
	\end{align*}
	Therefore, the stability condition $(C_2<0)$ is obtained by requiring that $\varpi^2>0$. 

	\subsection{Parameters space}
	Equations~\eqref{eq:en_par_norm} and~\eqref{eq:stability} can be used to track the set of parameters $\{x_0,\eta,y\}$ that describes stable $(\varpi^2>0)$ thin shells wormhole solutions supported by positive $(\tilde{\gamma}>0)$ or negative $(\tilde{\gamma}<0)$ surface energy densities. Let us investigate these two scenarios.
	
	\subsubsection*{Positive energy stable configurations}
	By requiring $\varpi^2>0$ and $\tilde{\gamma}>0$, one finds that the dimensionless parameters $\{x_0,\eta,y\}$ are constrained by	  
	\begin{align}
		\label{eq:x0_sp}&\dfrac{2}{15}(10-\sqrt{10})<x_0<\dfrac{2}{15}(10+\sqrt{10}) \\
		\label{eq:eta_sp}&\eta_1^- <\eta<\eta_1^+,\\
		\label{eq:y_sp}&y_1^-<y<y_1^+,
	\end{align} 
	where
	\begin{align*}
		\eta_1^\pm &= \frac{-15+34x_0-12x_0^2}{33-28x_0+6x_0^2}\\ &\pm 2\frac{\sqrt{-(216-504x_0+399x_0^2-130x_0^3+15x_0^4)}}{33-28x_0+6x_0^2},\numberthis\\
		y_1^- &= \dfrac{a-2\sqrt{b}}{\eta^2-2\eta+17},\numberthis\\
		y_1^+ &= \dfrac{4x_0(3-x_0)}{\eta+7},\numberthis
	\end{align*}
	with $a = 2x_0(17-\eta+(\eta-7)x_0)$ and $b = -x_0^2(17-2\eta+17\eta^2-2x_0(17+8\eta+7\eta^2)+x_0^2(19+3\eta(\eta + 2)))$. Equations~(\ref{eq:x0_sp})-(\ref{eq:y_sp}) determine the possible stable, positive energy (SPE) configurations allowed by gluing two RN spacetimes in Palatini $f(\mathcal{R})$ framework. The banana-shaped blue region in Fig.~\ref{fig:space_of_param} represents the parameter space of SPE wormholes.
	
	\subsubsection*{Negative energy stable configurations}
\begin{figure}
		\includegraphics[width=\linewidth]{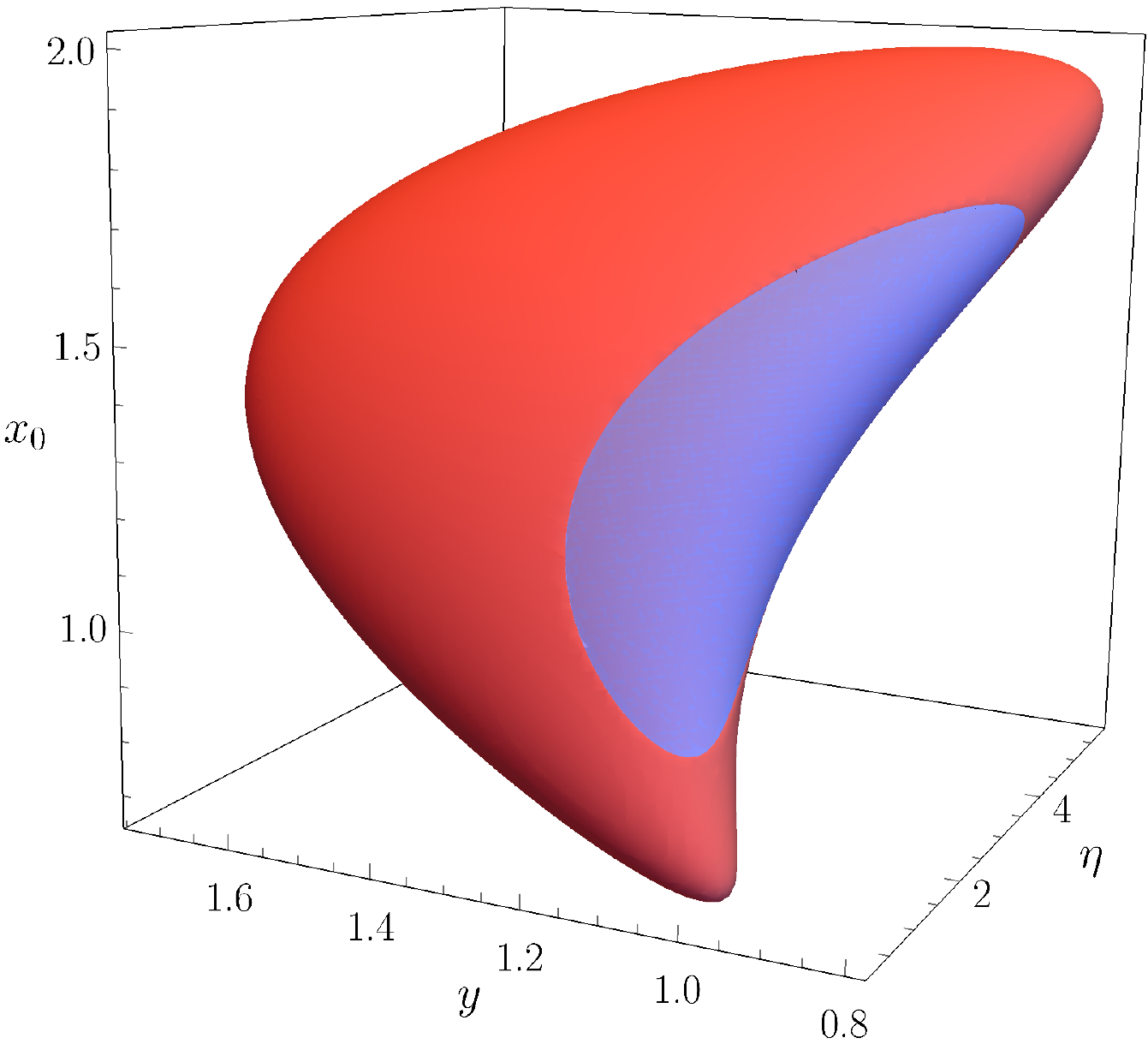}
		\caption{Parameter spaces for positive (blue region) and negative (red region) energy stable asymmetric wormholes.}
		\label{fig:space_of_param}
	\end{figure}

	Now, looking for $\varpi^2>0$ and $\gamma<0$, one finds stable, negative energy (SNE) wormhole configurations  --  that are associated with two different parameter spaces. The first group of solutions lies in the region identified by the constraints:
	\begin{align}
		\label{eq:sn_i}&\text{If }\dfrac{2}{3}<x_0\leq \dfrac{2}{15}(10-\sqrt{10}),\, \eta_2^-<\eta<\eta_2^{+},\\
		&\text{if }\dfrac{2}{15}(10-\sqrt{10})<x_0<1,\,\eta_2^-<\eta\leq\eta_1^-\text{ or }\eta_1^+<\eta<\eta_2^+,\\
		&\text{if }1\leq x_0<\dfrac{2}{15}(10+\sqrt{10}),\,\eta_2^-<\eta<\eta_1^-\text{ or }\eta_1^+<\eta<\eta_2^+,\\
		&\text{if }x_0 = \dfrac{2}{15}(10+\sqrt{10}),\, \eta_2^-<\eta<\eta_1^- \text{ or } \eta_1^-<\eta<\eta_2^+,\\
		\label{eq:sn_f}&\text{if }\dfrac{2}{15}(10+\sqrt{10})<x_0<2,\,\eta_2^-<\eta<\eta_2^+,
	\end{align}
	with dimensionless charge (in $\mathcal{M}_-$) bounded by
	\begin{equation}
		\label{eq:y_sn_1}
		y_1^-<y<y_2^+,
	\end{equation}
	where
	\begin{align*}
		\eta_2^{\pm} &= \dfrac{1+8x_0-3x_0^2}{x_0(3x_0-14)+17}\\
		&\pm\dfrac{2\sqrt{2}\sqrt{-(6x_0^4-40x_0^3+99x_0^2-104x_0+36)}}{x_0(3x_0-14)+17},\numberthis\\
		y_2^+ &= \dfrac{a+2\sqrt{b}}{\eta^2-2\eta+17}.
	\end{align*}
	The second group of solutions lies in the region identified by the constraints:
	\begin{align}
		\label{eq:sn_2_i}&\text{If }\dfrac{2}{15}(10-\sqrt{10})<x_0<1,\, \eta_1^-<\eta<\eta_1^+,\\
		&\text{if }1<x_0<\dfrac{2}{15}(10+\sqrt{10}),\, \eta_1^-\leq\eta\leq\eta_1^+,\\
		\label{eq:sn_2_f}&\text{if }x_0 = \dfrac{2}{15}(10+\sqrt{10}), \eta =  \eta_1^-,
	\end{align}
	with dimensionless charge constrained by
	\begin{equation}
		\label{eq:y_sn_2}
		y_1^+<y<y_2^+.
	\end{equation}
	The union of the two parameter spaces that identify SNE asymmetric wormholes is plotted in red in Fig.~\ref{fig:space_of_param}.

	As we can see in Fig.~\ref{fig:space_of_param} SPE space of parameters is embedded onto the SNE space of parameters. However it has to be pointed out that there is not any intersection between both regions. Then, as one can expect, there is not any configuration for which we can have positive and negative energy at the same time. In Appendix \ref{appendixOrtho} an orthographic projection of Fig.~\ref{fig:space_of_param} is depicted for the sake of showing in a more clear way the features explained above.
	
	Henceforth, we will consider only stable configurations, i.e, for now on we will always locate the throat of the wormhole at $x_0$. 
	
	\subsubsection*{Event horizon location}
	Since (two-way) traversable wormholes must have their two sides causally connected, an important point to set up is whether a RN-AWH has an event horizon, and if it does, whether the throat covers it. Using the dimensionless variables, one finds that the event horizon location, on each side, is given by
	\begin{align}
		\label{eq:ev_-} x_{h-} &= 1+\sqrt{1-y},\\
		\label{eq:ev_+} x_{h+} &= \xi+\sqrt{\xi^2-\eta y}.
	\end{align}
	Then, in order to prevent the two universes to be causally disconnected, $x_0$ must be put above $x_{h\pm}$ on each side of the wormhole. This requirement leads to the following constraints:
	\begin{align*}
		&0<y\leq 1 \text{ and } x_0> 1+\sqrt{1-y},\numberthis\\
		&y> 1 \text{ and } 0<x_0<\dfrac{y}{2} \text{ or } x_0>\dfrac{y}{2},\numberthis\\ &0<\eta\leq 1+ \dfrac{2x_0(x_0-1-\sqrt{x_0(x_0-2)+y})}{y}. \numberthis\\ 
	\end{align*}
	
	In the subsequent sections, we investigate only stable two-way traversable wormholes, that is either a RN black hole glued with a RN naked singularity or two naked singularities glued together. The possibility of having two RN black holes glued together is excluded because it leads to unstable configurations. Note that, when we refer to either RN black hole or RN naked singularity, we are referring to their space-times beyond the event horizon and the singularity, respectively.

	\section{Absorption and spectral lines}\label{sec:abs}
	\subsection{Wave equation}
	Let us consider a massless scalar field, $\Phi$, lying in a RN-AWH background. The dynamics of this field, on each side of the wormhole, is described by the Klein-Gordon equation
	\begin{equation}
		\label{eq:kg}
		\Box_{\pm} \Phi_\pm = 0,
	\end{equation}
	where $\Box_\pm$ and $\Phi_\pm$ denote the d'Alembertian operator and the scalar field, respectively, on each side of the throat. Due to spherical symmetry, the solution of Eq.~\eqref{eq:kg} can be written as
	\begin{equation}
		\label{eq:field_decomp}
		\Phi_\pm = \dfrac{\psi_{\pm}(x_\pm)}{x_\pm}Y_{\ell m}(\theta,\phi)e^{-i \tilde{\omega} \tilde{t}},
	\end{equation}
	where $\tilde{\omega}$ is a dimensionless frequency (defined by $\tilde{\omega} \equiv \omega M_-$), and the radial functions $\psi_\pm$ satisfy
	\begin{equation}
		\label{eq:field}
		f_{\pm}(x_\pm)\dfrac{d}{dx_\pm}\left(f_{\pm}(x_\pm)\dfrac{d\psi_\pm}{dx_\pm}\right)+\left(\tilde{\omega}^2-\tilde{V}_\pm(x_\pm)\right)\psi_\pm = 0,
	\end{equation} 
	with $\tilde{V}_\pm$ being the dimensionless effective potential on each side of the throat, given by
	\begin{equation}
		\label{eq:effective_potential}
		\tilde{V}_\pm(x_\pm) = \dfrac{f_\pm(x_\pm)}{x_\pm}\dfrac{df_\pm}{dx_\pm} + \dfrac{f_\pm(x_\pm)}{x_\pm^2}\ell(\ell+1).
	\end{equation}
The metric functions $f_\pm(x_\pm)$ explicitly written in terms of the dimensionless radial coordinates are
	\begin{align}
		\label{eq:f_+x} f_-(x_-) &= 1 - \dfrac{2}{x_-}+\dfrac{y}{x_-^2},\\
		\label{eq:f_-x} f_+(x_+) &= 1 - \dfrac{2\xi}{x_+}+\dfrac{\eta y}{x_+^2}.
	\end{align}
One could think that the non-differentiability of the metric would introduce a delta-type contribution at the throat in the effective potential. However, one can argue that, since the metric is continuous across the shell and the d'Alembertian operator contributes only with the first derivative of the metric, $\partial_\mu g_{\alpha\beta}$, and the first derivative of the metric determinant, $\partial_\mu g = g g^{\alpha\beta}\partial_{\mu}g_{\alpha\beta}$; no delta-type distribution will appear in the effective potential. In the distributional approach, we have~\cite{clarke:1987}
\begin{align}
\label{eq:deriv_metric_distr}
\partial_{\mu}\underline{g}_{\alpha\beta} &= \partial_{\mu}g^{+}_{\alpha\beta}\,\underline{\theta}+\partial_{\mu}g^{-}_{\alpha\beta}\,(\underline{1}-\underline{\theta})+n_{\mu}[g_{\alpha\beta}]\,\underline{\delta}^{\Sigma},
\end{align}	
and analogously
\begin{align}
\partial_{\mu}\underline{g} &= \partial_{\mu}g^{+}\,\underline{\theta}+\partial_{\mu}g^{-}\,(\underline{1}-\underline{\theta})+n_{\mu}[g]\,\underline{\delta}^{\Sigma}.
\end{align}
If the metric is continuous across the shell, both $[g_{\alpha\beta}]$ and $[g]$ must vanish, therefore no delta-type distribution appears in the effective potential. However, if the metric is discontinuous across the shell, as in the case of dirty black holes~\cite{leung:1999}, one expects the appearance of a delta-type contribution in the effective potential.

	It will be convenient to introduce a global radial coordinate to describe the spacetime, which is implicitly defined by
	\begin{equation}
		\label{eq:global_radial_coord}
		d\xst = \pm \dfrac{dx_{\pm}}{f_{\pm}(x_{\pm})}.
	\end{equation}
	The main advantage of this new coordinate is that it combines the information of two independent domains, namely $x_-\in [x_0,\infty)$ and $x_+\in [x_0,\infty)$, in a single domain $\xst\in (-\infty,\infty)$. Moreover, with a suitable choice of integration constant, the throat location moves to $\xst(x_0)=0$. 
	By using the global radial coordinate~\eqref{eq:global_radial_coord}, one may write Eq.~\eqref{eq:field} as a Schr{\"o}dinger-like equation, namely
	\begin{equation}
		\label{eq:field_xst}
		\dfrac{d^2\psi}{d\xst^2} + \left(\tilde{\omega}^2-\tilde{V}(\xst)\right)\psi = 0,
	\end{equation}
	where we dropped the $\pm$ in the subscripts, since the global radial coordinate allows us to express the radial function and the effective potential as functions of $\xst$, respectively, $\psi(\xst)$ and $\tilde{V}(\xst)$. 

	\subsection{Effective potential}
\begin{figure}
		\includegraphics[width=\linewidth]{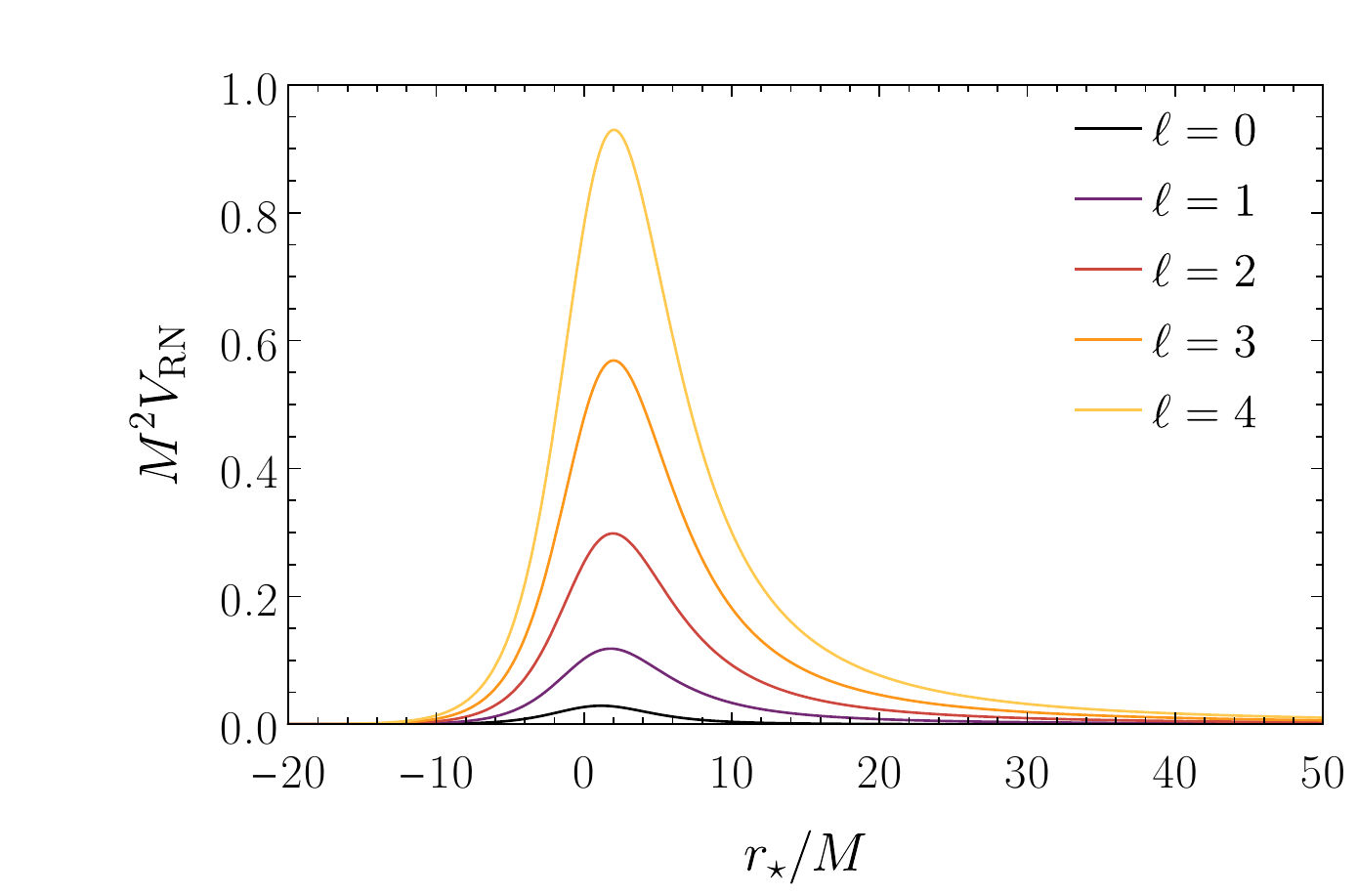}
		\caption{Effective potential of a RN black hole with $Q^2/M^2=0.5$.}
		\label{fig:effective_potential_RN}
	\end{figure}

	The effective potential plays a key role in understanding the dynamics of the scalar field. Since the RN-AWH consists of two RN spacetimes glued, it is convenient to analyze the effective potential of the RN spacetime first. The effective potential of a RN spacetime is~\cite{CDO}
	
	\begin{equation}
		\label{eq:effective_potential_RN}
		V_\text{RN}(r) = \dfrac{f(r)}{r}\dfrac{df}{dr} + \dfrac{f(r)}{r^2}\ell(\ell+1),
	\end{equation}
	where $f(r) = 1-2M/r+Q^2/r^2$, with $M$ and $Q$ being the mass and charge of the black hole, respectively. Similar to Eq.~\eqref{eq:global_radial_coord}, one may define a new radial coordinate, the so-called tortoise coordinate, that moves the event horizon location to $-\infty$, namely $dr_\star = dr/f(r)$, so that the causally connected part of the manifold is described by $r_\star\in(-\infty,\infty)$. From Eq.~\eqref{eq:effective_potential_RN} we notice that the effective potential vanishes at the event horizon $r=M+\sqrt{M^2-Q^2}$ and at the spacial infinity, i.e, $V_\text{RN}\to 0$, $r_{\star}\to\pm\infty$. In Fig.~\ref{fig:effective_potential_RN} we plot the effective potential, for $Q^2/M^2=0.5$ and some angular momentum numbers $\ell$, as a function of the tortoise coordinate. As can be seen from Fig.~\ref{fig:effective_potential_RN}, the effective potential has a peak that varies with $\ell$.
	In the eikonal limit $(\ell \gg 1)$ the dominant term of the effective potential is proportional to $f(r)/r^2$, i.e, it has the same dependence on $r$ as the classical scattering potential that appears when studying the motion of null-like particles in the RN background. Consequently, for large values of $\ell$, the location of the effective potential peak is at the photon sphere, namely
	\begin{equation}
		\label{eq:photon_orbits}
		r_\gamma = \dfrac{3M}{2}+\dfrac{1}{2}\sqrt{9M^2-8Q^2}.
	\end{equation}
	In the classical scattering process, the peak of the effective potential has the value $V_{\text{RN}}(r_\gamma)=f(r_\gamma)/r_\gamma^2 = 1/b_c^2$, where $b_c$ is the so-called critical impact parameter. 
	
\begin{figure*}
		\centering
		\includegraphics[width=\columnwidth]{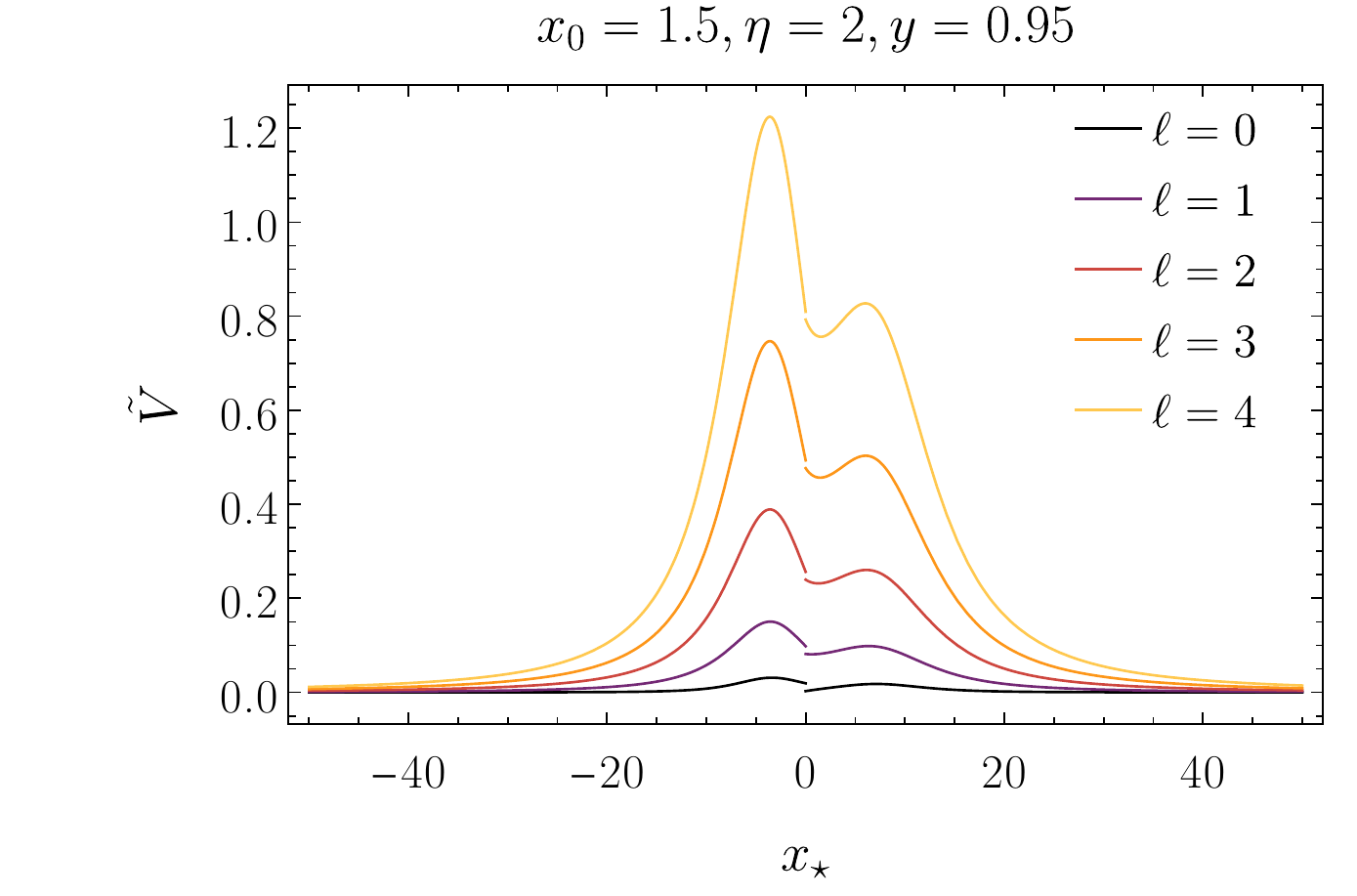} \includegraphics[width=\columnwidth]{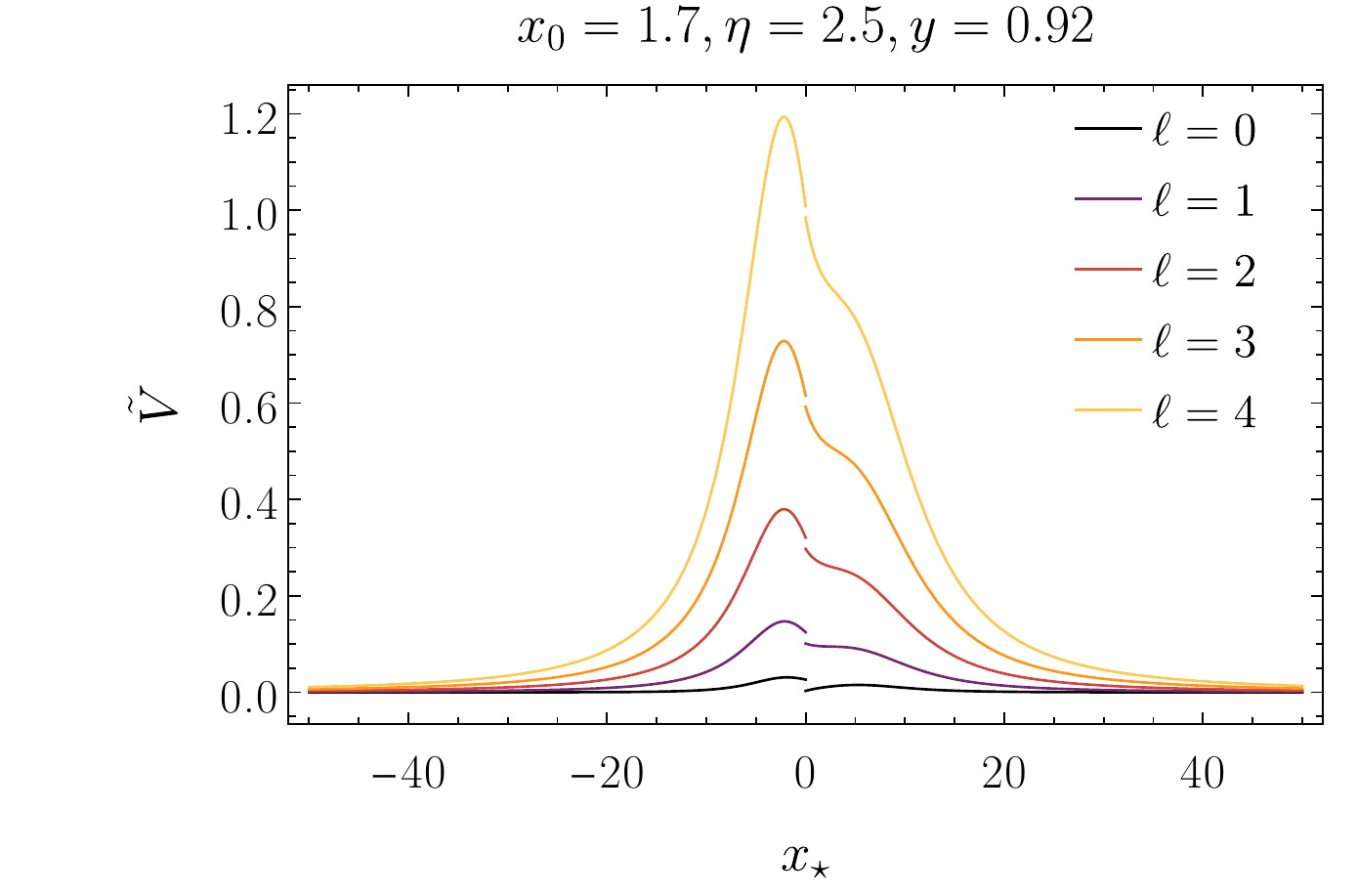}

		\includegraphics[width=\columnwidth]{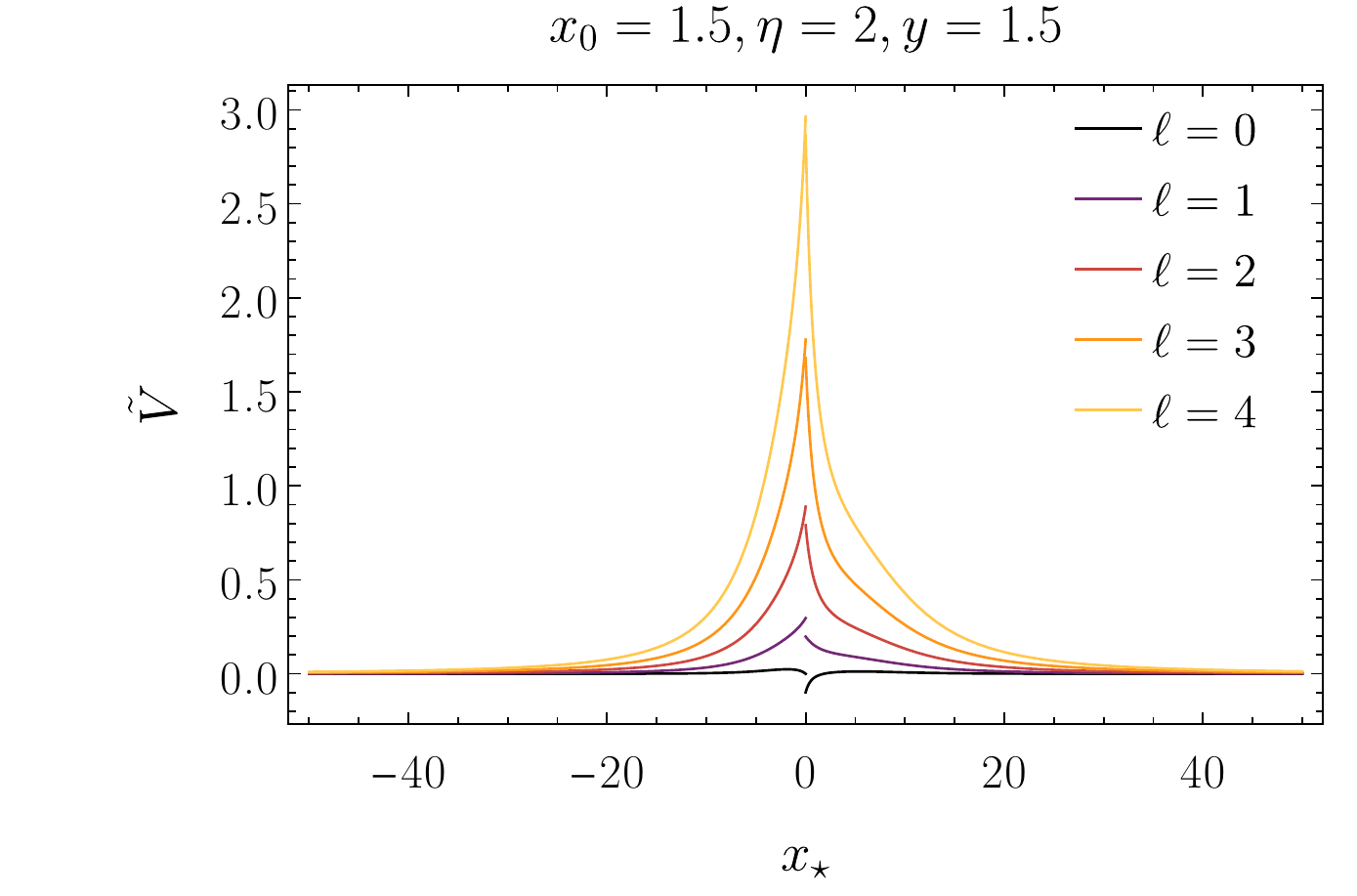}\includegraphics[width=\columnwidth]{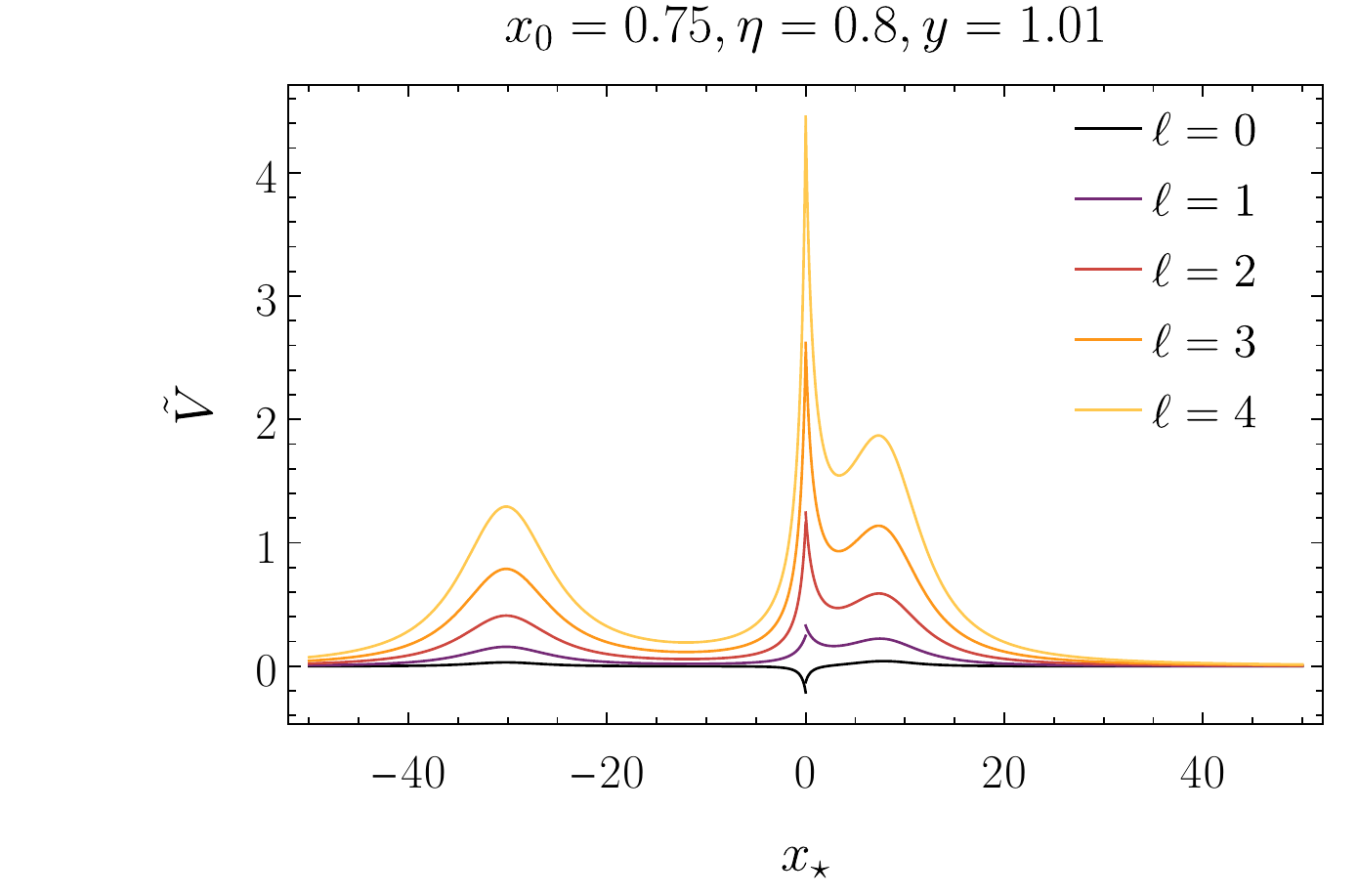}
		\caption{Typical behaviors of the effective potential of RN-AWHs. In the top row we plot the effective potential for two SPE configurations and in the bottom row we plot the effective potential for two SNE configurations.}
		\label{fig:effective_potential_RNAWH}
	\end{figure*}

\begin{figure*}
		\centering
		\includegraphics[width=\columnwidth]{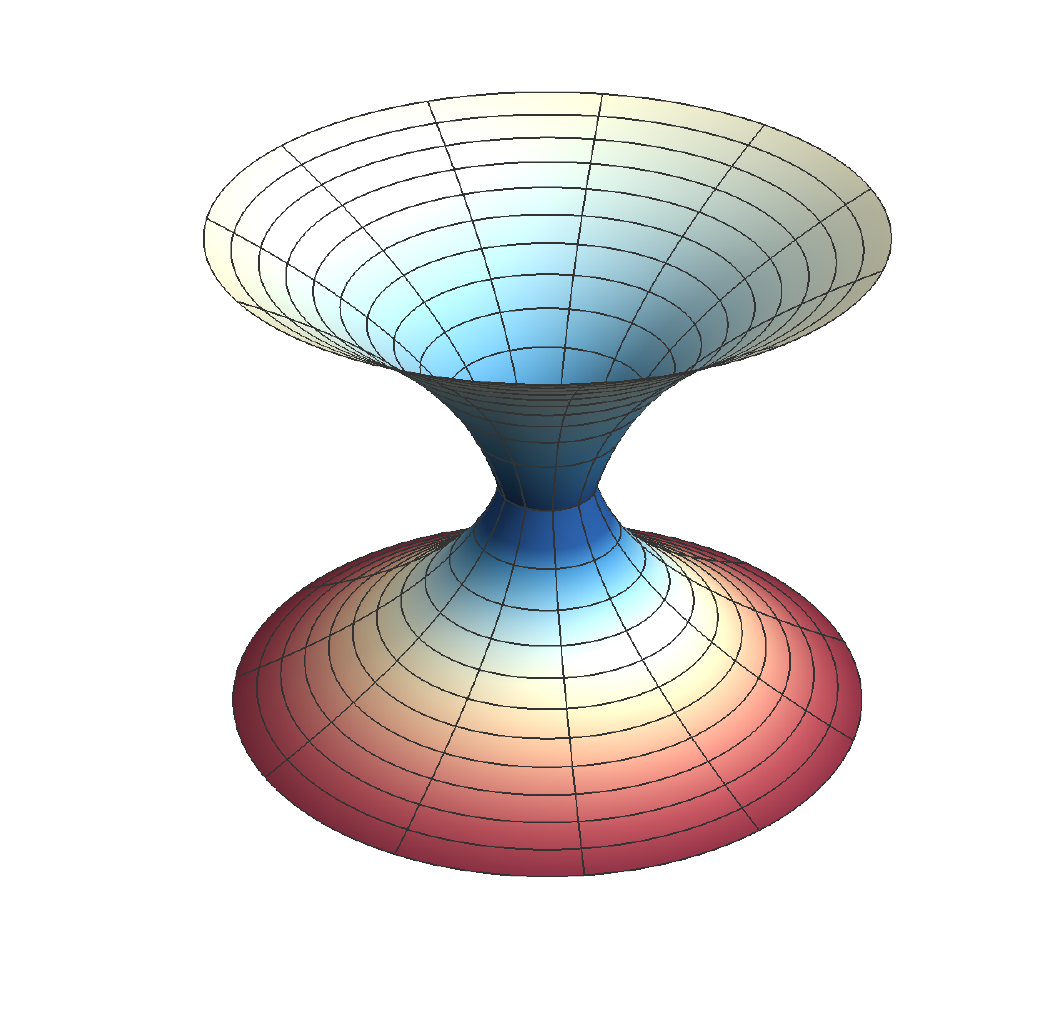} \includegraphics[width=\columnwidth]{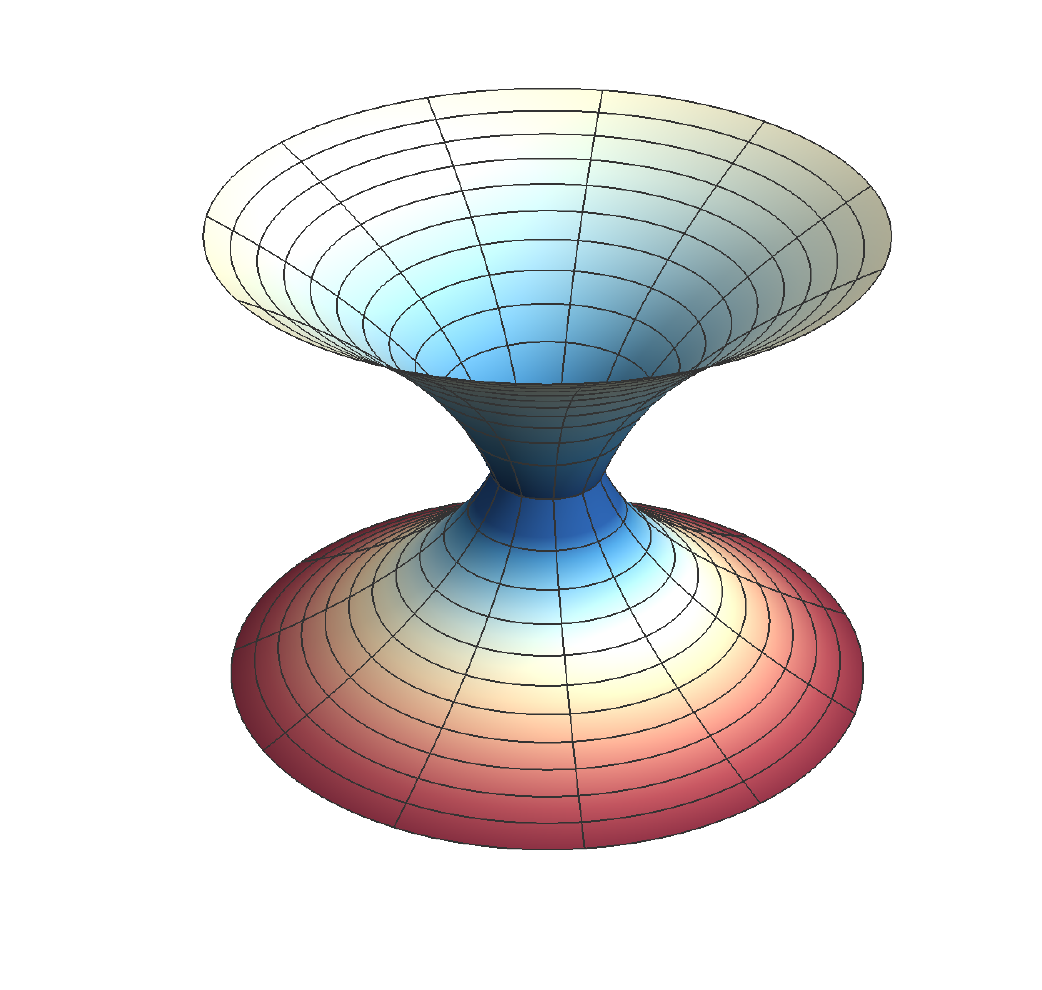}

		\includegraphics[width=\columnwidth]{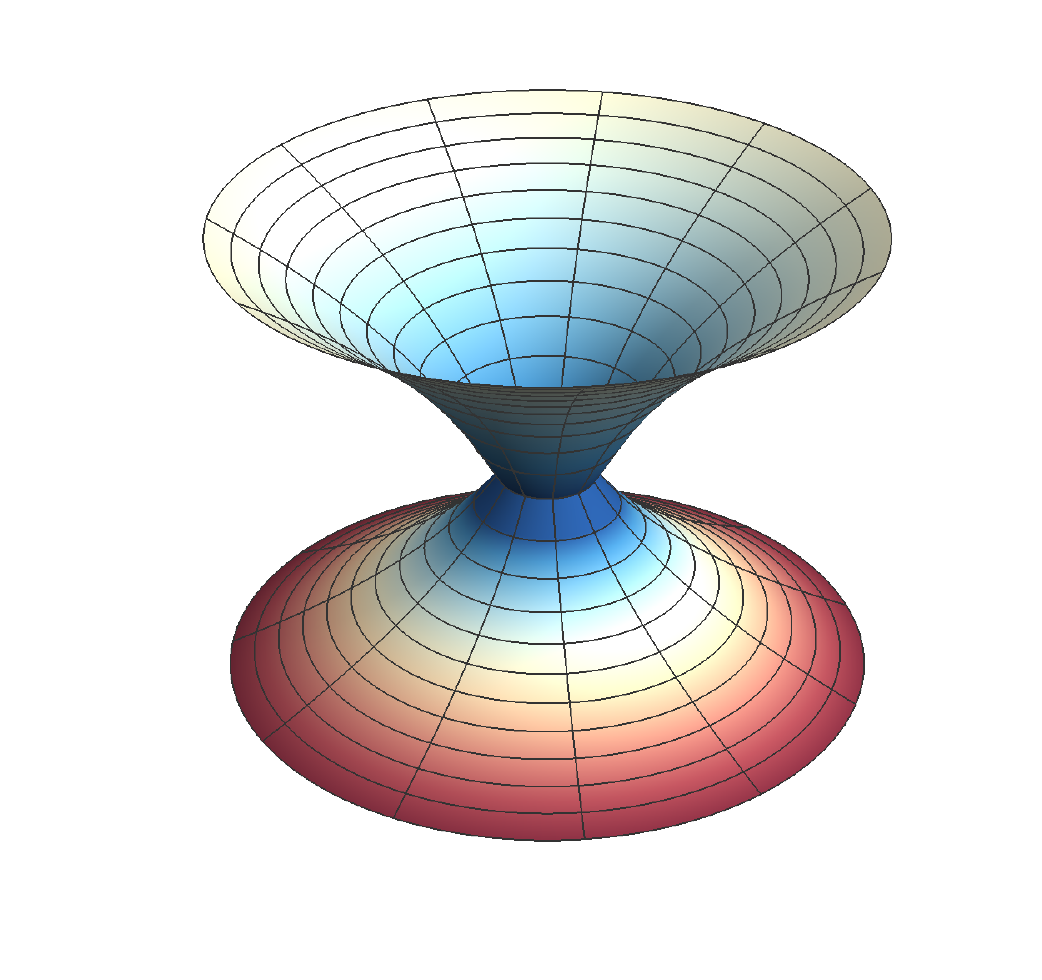}\includegraphics[width=\columnwidth]{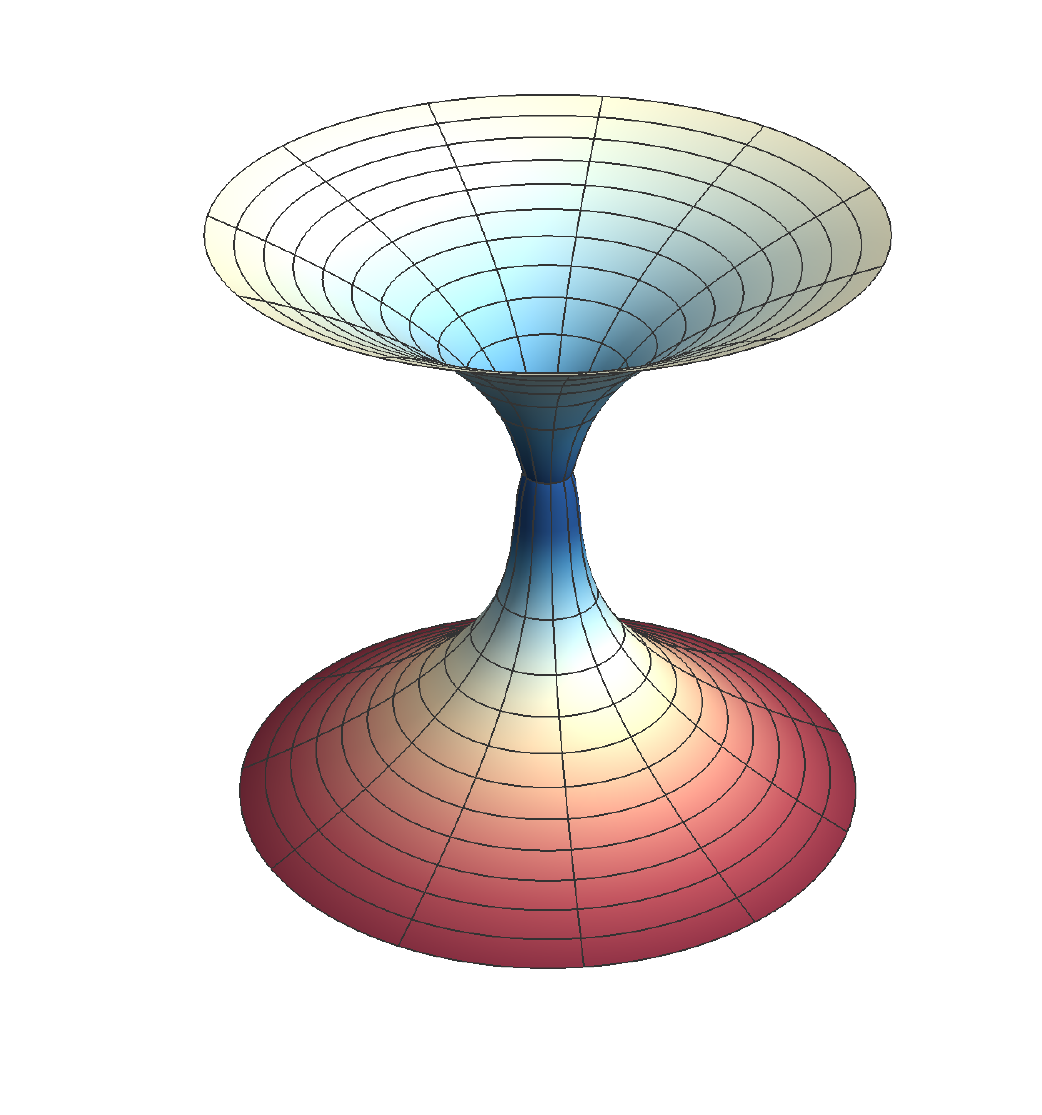}
		\caption{Embedding diagrams of the RN-AWHs considered in Fig.~\ref{fig:effective_potential_RNAWH} (the display order is the same as in Fig.~\ref{fig:effective_potential_RNAWH}).}
		\label{fig:embedding_diagrams}
	\end{figure*}

	Now we can discuss the effective potential of RN-AWHs. Just like in the black hole case, the effective potential of the RN-AWH vanishes far from the throat, i.e, $\tilde{V}\to 0$, $\xst\to\pm\infty$. As we get closer to the throat, the effective potential increases and it may have a peak on each side of the throat, depending on the shell location. Since the metric function is not differentiable at $\xst=0$, the effective potential may have a discontinuity (``jump'') at the throat, that is, $[\tilde{V}]\neq 0$, when gluing different spacetimes. In Fig.~\ref{fig:effective_potential_RNAWH} we plot some typical behaviors of the effective potential for RN-AWHs. We notice that the number of peaks varies, depending on the throat location, since it may be located before or after the peak of the effective potential on each side. By gluing a RN black hole with a RN naked singularity at least one peak is present. By gluing two different RN naked singularities a sharp \textit{discontinuous peak} appears apart from the possible smooth peaks; however, it is important to point out that, at this peak, $d\tilde{V}/d\xst = 0$ is not satisfied. Actually, $\tilde{V}$ is not differentiable at $\xst = 0$ (one can also find discontinuous effective potentials in Refs.~\cite{leung:1999,barausse:2014,macedo:2016}). In Fig.~\ref{fig:embedding_diagrams}, we exhibit the embedding diagrams of the RN-AWHs considered in Fig.~\ref{fig:effective_potential_RNAWH}.
	
	\subsection{Boundary conditions}
\begin{figure}
		\begin{center}
			\begin{tikzpicture}[scale=0.89, transform shape]
				\path 
				+(90:4)  coordinate[label=90:$i^+$]  (IItop)
				+(-90:4) coordinate[label=-90:$i^-$] (IIbot)
				+(0:4)   coordinate[label=0:$i^0_{{\cal M}_+}$] (IIright)
				+(180:4) coordinate[label=180:$i^0_{{\cal M}_-}$] (IIleft)
				;
				\draw (IIleft) -- 
				node[midway, above left]    {$\mathscr{I}^+_{{\cal M}_-}$}
				(IItop) --
				node[midway, above right]    {$\mathscr{I}^+_{{\cal M}_+}$}
				(IIright) -- 
				node[midway, below right] {$\mathscr{I}^-_{{\cal M}_+}$}
				(IIbot) --
				node[midway, below left] {$\mathscr{I}^-_{{\cal M}_-}$}
				(IIleft) -- cycle
				;
				\draw[densely dotted] (IItop) -- 
				(IIbot) -- cycle
				node[near start, above left, sloped]    {$\xst=0$}
				;
				\draw [->] (0.5,0.5) --  (1.5,1.5);
				\draw [->] (0.5+0.176777,0.5-0.176777) --  (1.5+0.176777,1.5-0.176777);
				\draw [->] (-0.5,0.5) --  (-1.5,1.5);
				\draw [->] (1.5,-1.5) --  (0.5,-0.5);
				\draw [->] (1.5-0.176777,-1.5-0.176777) --  (0.5-0.176777,-0.5-0.176777);
				\draw [->] (1.5+0.176777,-1.5+0.176777) --  (0.5+0.176777,-0.5+0.176777);
			\end{tikzpicture}
		\end{center}
		\caption{Carter-Penrose diagram of a RN-AWH. Each triangle represents an asymptotically flat spacetime, connected by a wormhole throat, represented by the vertical dotted line. The arrows illustrate the scattering process. Here, $\mathscr{I}^+$ and $\mathscr{I}^-$ represent, respectively, the future and past null infinities, $i^0$ indicates the space-like infinity and $i^+$ and $i^-$ are, respectively, the future and past time-like infinities. 
(Labels without subscript represent regions of the two sides of the wormhole that are superposed in the Carter-Penrose diagram.)
		}
		\label{fig:penrose_diag_scatt}
	\end{figure}
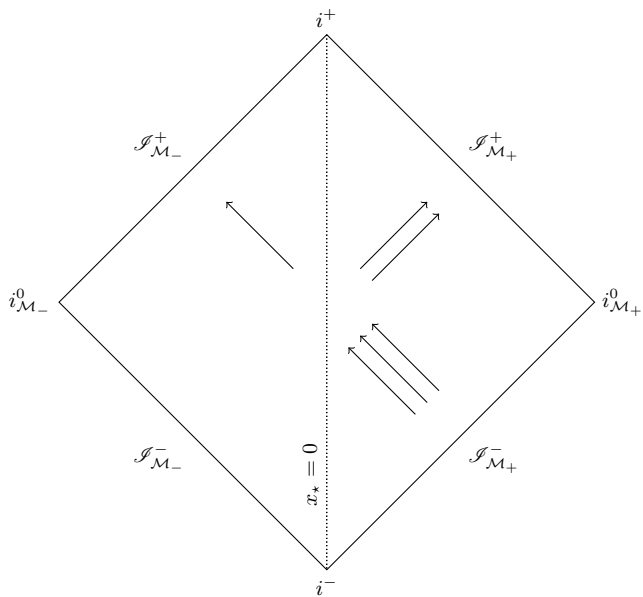

	In Fig.~\ref{fig:penrose_diag_scatt}, using the Carter-Penrose diagram of RN-AWHs, we illustrate the scattering process. 
Let us consider a monochromatic plane scalar wave incoming from the past null infinity of ${\cal M}_+$, $\mathscr{I}^-_{{\cal M}_+}$. This wave will interact with the effective potential; part of it will be reflected to the future null infinity of ${\cal M}_+$, $\mathscr{I}^+_{{\cal M}_+}$; and part of it will be transmitted to the future null infinity of ${\cal M}_-$, $\mathscr{I}^+_{{\cal M}_-}$. Therefore, the stationary boundary conditions of this phenomenon consist of a composition of ingoing and outgoing (\textit{distorted}) plane waves far from the object in one side of the wormhole, and purely outgoing waves far from the object on the other side, i.e,
	\begin{equation}
		\psi(\xst)\sim\left\{
		\begin{array}{ll}
			e^{-i \tilde{\omega} \xst}+{\cal R}_{\tilde{\omega} \ell }e^{i \tilde{\omega} \xst},&\xst\to+\infty, \\
			{\cal T}_{\tilde{\omega} \ell }e^{-i\tilde{\omega} \xst},& \xst\to-\infty,
		\end{array}\right.\label{eq:inmodes}
	\end{equation}
	where ${\cal R}_{\tilde{\omega} \ell }$ and ${\cal T}_{\tilde{\omega} \ell }$ are complex coefficients related to the reflection and transmission coefficients, respectively.	
In order to obtain ${\cal R}_{\tilde{\omega} \ell }$ and ${\cal T}_{\tilde{\omega} \ell }$, one performs an integration from one asymptotic region to the 
other. 

We have also to  
specify the behavior of the field at the throat. Just like the metric, we assume that the field is continuous across the throat, i.e, $[\Phi]=0$ (hence $[\psi]=0$), and since no delta-type distribution appears in the effective potential, we can assume that the field is also differentiable at the shell, that is, $[d\psi/d\xst]= 0$ (see Refs.~\cite{aneesh:2018,cardoso:2016} for some works in the literature considering the differentiability of the field, despite the non differentiability of the metric function).
	
	\subsection{Scalar absorption}
	By using the partial wave expansion together with the boundary conditions~\eqref{eq:inmodes}, one can write the (dimensionless) total scalar absorption cross section of RN and RN-AWHs as~\cite{CDO,limajr:2020}
	\begin{equation}
		\label{eq:abs_total_cross_section} \tilde{\sigma}_{\text{abs}}=\sum_{\ell = 0}^{\infty}\tilde{\sigma}_\ell,
	\end{equation}
	where $\tilde{\sigma}_\ell \equiv\pi(2\ell +1)\Gamma_{\tilde{\omega}\ell}/\tilde{\omega}^2$ are the (dimensionless) partial absorption cross sections, and $\Gamma_{\tilde{\omega}\ell}\equiv 1-|{\cal R}_{\tilde{\omega}\ell}|^2= |{\cal T}_{\tilde{\omega}\ell}|^2$ is the so-called grey-body factor, i.e., the transmission probability of a mode with frequency $\tilde{\omega}$~\cite{K:2004}. In black hole scenarios, the total absorption cross section~\eqref{eq:abs_total_cross_section} has two well-known limits for stationary geometries, namely, the low--frequency regime $(\tilde{\omega}\to 0)$, where the absorption cross section goes to the area of the black hole~\cite{higuchi:2001}, and the high-frequency regime $(\tilde{\omega}\to\infty)$, where the absorption cross section oscillates around the geometrical absorption cross section~\cite{decanini:2011} (the area of a disk with radius equal to the critical impact parameter). In wormhole scenarios these limits are not so clear. 
For instance, some previous results in the literature show that, in the zero-frequency regime, the total absorption cross section for wormholes can differ from black hole cases~\cite{delhom:2019,limajr:2020}. 
	
\begin{figure}
		\centering
		\includegraphics[width=\columnwidth]{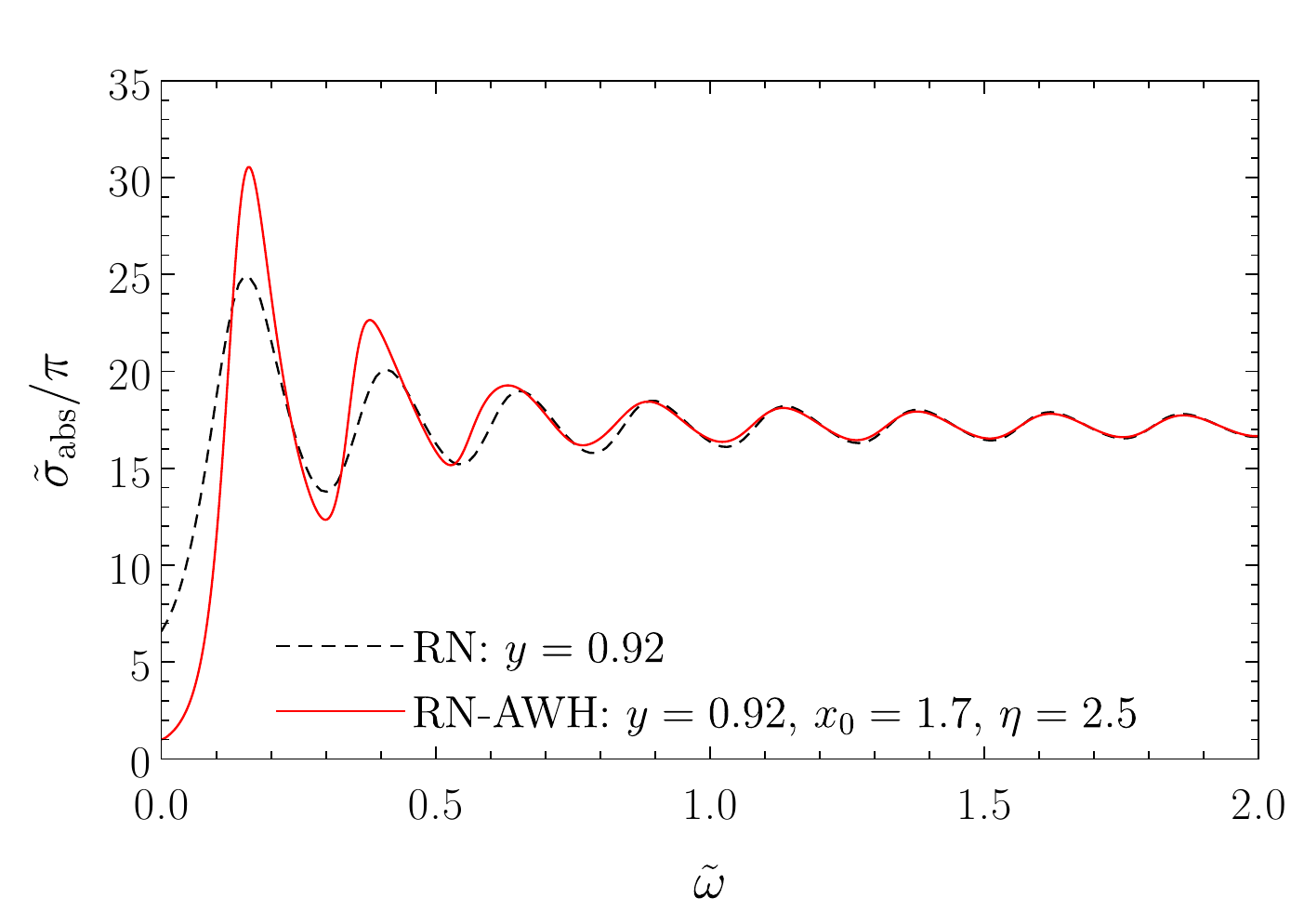}\hspace{0.3cm} \includegraphics[width=\columnwidth]{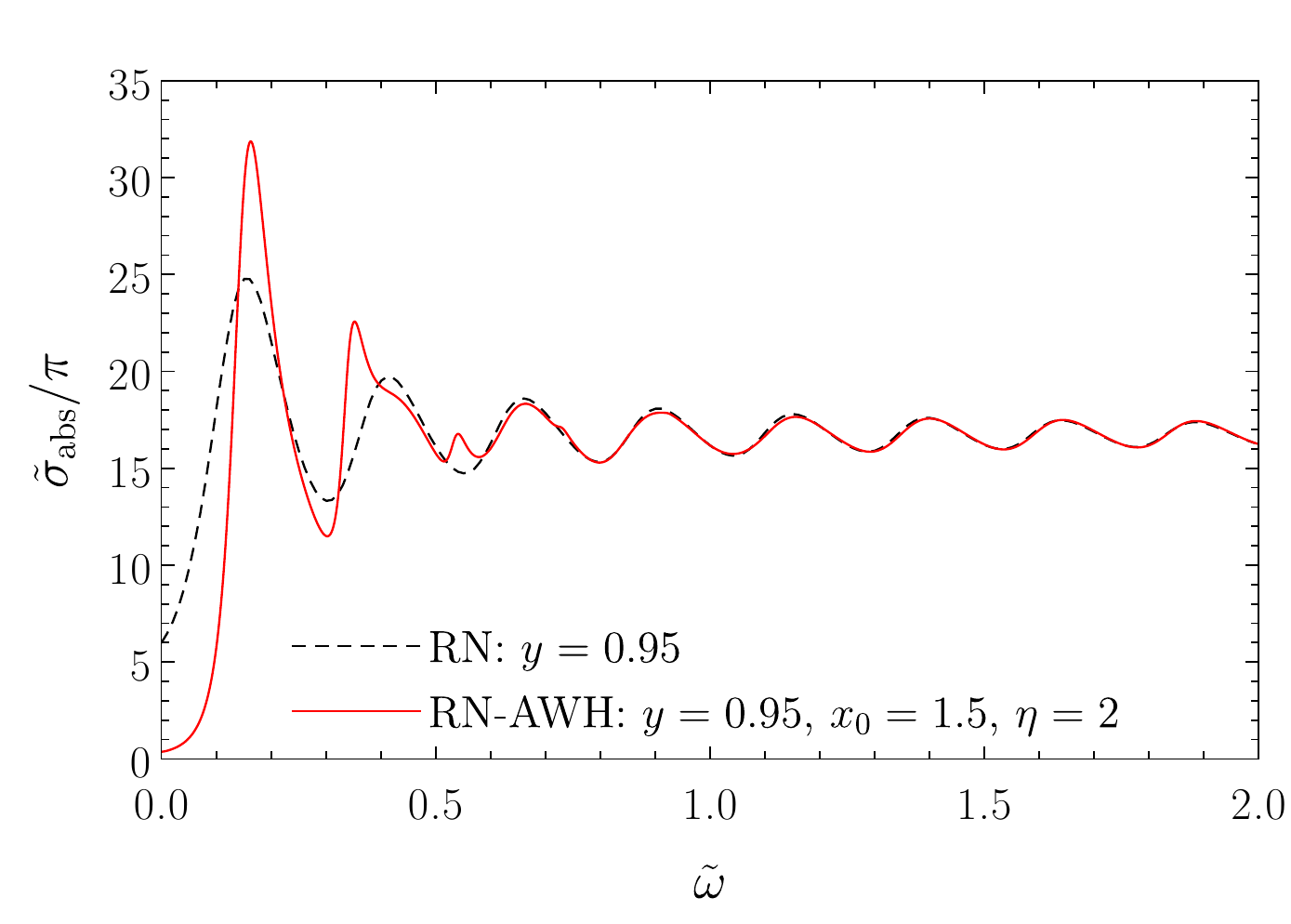}
		\caption{Comparison of the absorption cross section of RN-AWHs and RN black holes with the same charge-to-mass ratio.}
		\label{fig:comparison_RNAWH_RN}
	\end{figure}

	In Fig.~\ref{fig:comparison_RNAWH_RN} we plot the total absorption cross section of two SPE wormholes, namely $\{x_0=1.7,\eta=2.5,y=0.92\}$ and $\{x_0=1.5,\eta=2,y=0.95\}$, and compare it with the total absorption cross section of a RN black hole with the same charge-to-mass ratio as for the case of ${\cal M}_-$, i.e., $y=Q^{2}_-/M^2_-=Q^2/M^2$. The effective potential of the two wormhole configurations considered in Fig.~\ref{fig:comparison_RNAWH_RN} are exhibited in Fig.~\ref{fig:effective_potential_RNAWH}. We notice that, in the high-frequency regime, the total absorption cross section of RN-AWHs (constructed with a RN black hole in ${\cal M}_-$) goes to the RN black hole profile, i.e., it oscillates around the classical absorption cross section. However, at low-frequencies the behavior of the absorption spectra is different from the corresponding black hole case.
In the zero-frequency limit, the total absorption cross section of RN-AWHs is much smaller than the corresponding black hole one, in accordance to what has been obtained in other wormhole configurations~\cite{delhom:2019,limajr:2020}.

	In order to investigate how the energy density of the shell that supports the wormhole configuration affects the absorption spectra, let us first analyze SPE and SNE configurations with the same wormhole throat, $x_0$, and charge-to-charge ratio, $\eta$, but with dimensionless charge, $y$, constrained in different ways.
	By fixing $x_0=1.5$ and $\eta=2$, the energy parameter of the shell diminishes as one increases the charge-to-mass ratio $y$ (see Eq.~\eqref{eq:en_par_norm}). For these parameters, one finds that shells with positive-energy density can support stable solutions with dimensionless charge $0.919184<y<1$. However, to support stable wormholes with higher values of dimensionless charge, shells with negative energy content are required. With this choice of parameters, a negative-energy shell can support a stable wormhole with dimensionless charge $1<y<1.72787$.
	A glance at Fig.~\ref{fig:effective_potential_RNAWH} shows that these two families of configurations have solutions with significant differences in their effective potentials. For instance, we can have SPE configurations with $x_0=1.5$ and $\eta=2$, presenting two local maxima and a discontinuous valley between them for moderate-to-high values of $\ell$, while we can have SNE configurations, with $x_0=1.5$ and $\eta=2$, without a local maximum (where $d\tilde{V}/d\xst = 0$), but with a sharp peak near the shell for moderate-to-high values of $\ell$. One may expect that these differences in the effective potential lead to different behaviors of the absorption profile. This will be explored later in this section. 
	
\begin{figure}
		\centering
		\includegraphics[width=\columnwidth]{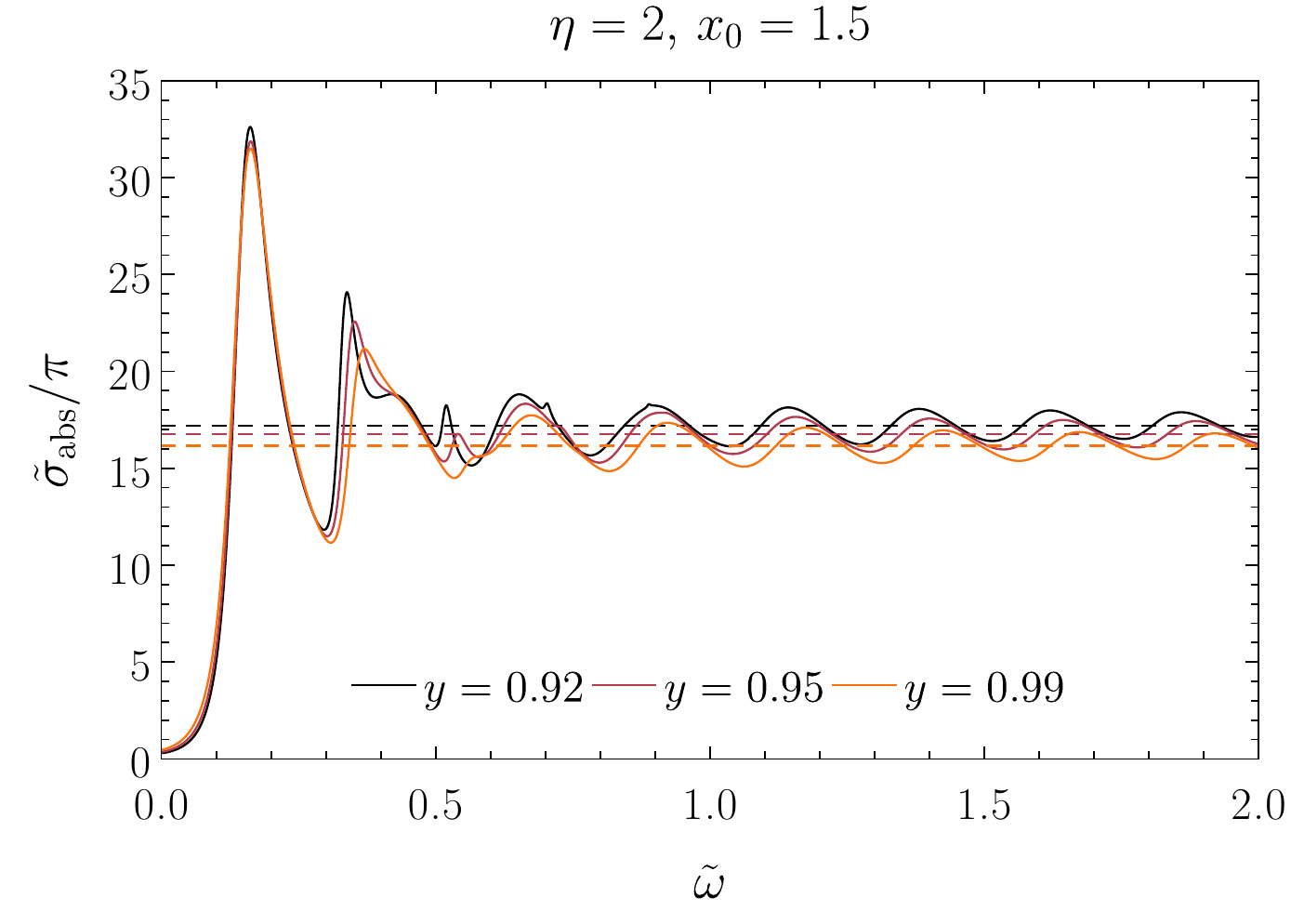}\hspace{0.3cm} \includegraphics[width=\columnwidth]{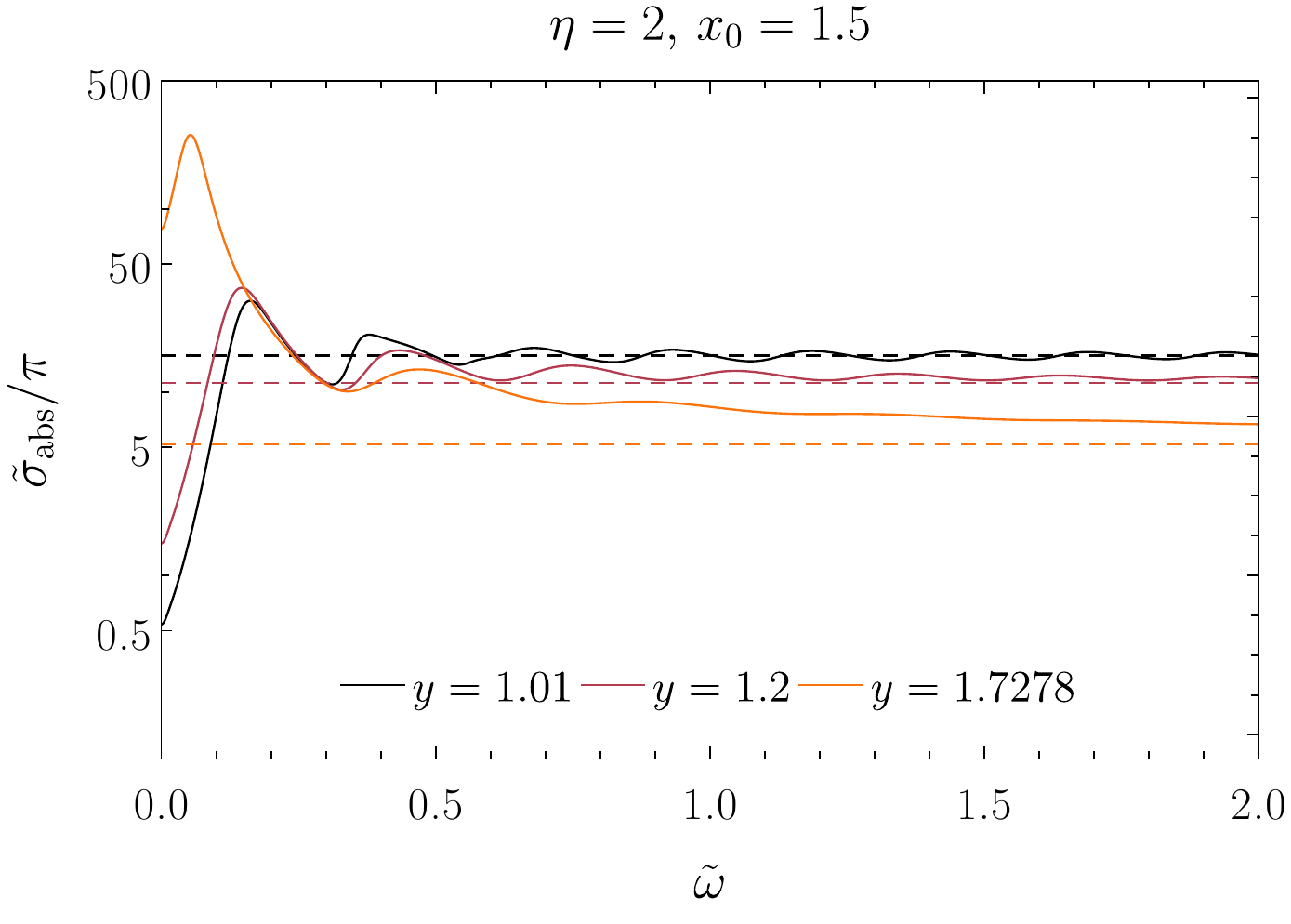}
		\caption{Absorption spectra of RN-AWHs with $x_0=1.5$ and $\eta=2$, supported by SPE- (top panel) and SNE (bottom panel) thin shells. 
		The dashed lines are the shadow areas associated with the highest peak of the effective potential in the eikonal limit $(\ell\gg 1)$. }
		\label{fig:absorption_spectra_RNAWH}
	\end{figure}

	In Fig.~\ref{fig:absorption_spectra_RNAWH}, we exhibit the absorption cross section of some SPE (top panel) and SNE (bottom panel) RN-AWHs configurations. We notice that increasing the dimensionless charge in ${\cal M}_-$ diminishes the total absorption cross section for moderate-to-high frequencies for both SPE and SNE wormholes. For SPE configurations, as we decrease $y$ (consequently increasing the energy parameter of the shell $\tilde{\gamma}$) additional peaks arise. These new peaks get higher and are shifted to the left as the positive-energy density of the shell increases. The new peaks are related with quasibound states that can exist around the throat~\cite{delhom:2019,limajr:2020}, that appear due to the presence of valleys in the effective potential (a discussion about them will follow in Sec.~\ref{sec:qbs}).  On the other hand, for SNE configurations, as $y$ increases (hence requiring ``more'' negative shells)
the behavior of the total absorption cross section can differ significantly from previous absorption profiles found in the literature. For this family of parameters, in the zero-frequency limit, as the energy density of the shell becomes more negative, the absorption cross section increases, getting bigger than the ones of standard black holes. This result should be related with the effective potential for $\ell=0$. 
Additionally, the oscillatory pattern of the total absorption cross section slowly diminishes and approaches a straight line. This high-frequency behavior is related with the sharp peaks that appear in the effective potential at the throat. In the eikonal limit, the total absorption cross section of spherically symmetric black holes oscillates around their shadow area, with the shadow radius being the critical impact parameter, which is associated with null geodesics trapped in the last photon orbit -- the photon sphere. In wormholes scenarios, it is possible to have an effective photon sphere at the throat, and it may cast novel shadows and different gravitational lensing features compared with black holes~\cite{wang:2020,shaikh:2019,Olmo:2023lil}. Here we notice that, the throat, where we have the sharp peak of the effective potential, acts like an effective photon sphere, which can be associated with an effective critical impact parameter, $b_{\text{ec}}=x_{0}/\sqrt{f_{\pm}(x_0)}$, that gets smaller as one increases the dimensionless charge $y$ (i.e. the sharp peak gets higher as $y$ approaches $y\approx 1.72787$). Hence, one expects that in the high-frequency regime, the absorption cross section goes to the area of this novel shadow, $A_{\text{ns}}=\pi x_{0}^2/f_{\pm}(x_0)$.

\begin{figure}
		\centering
		\includegraphics[width=\columnwidth]{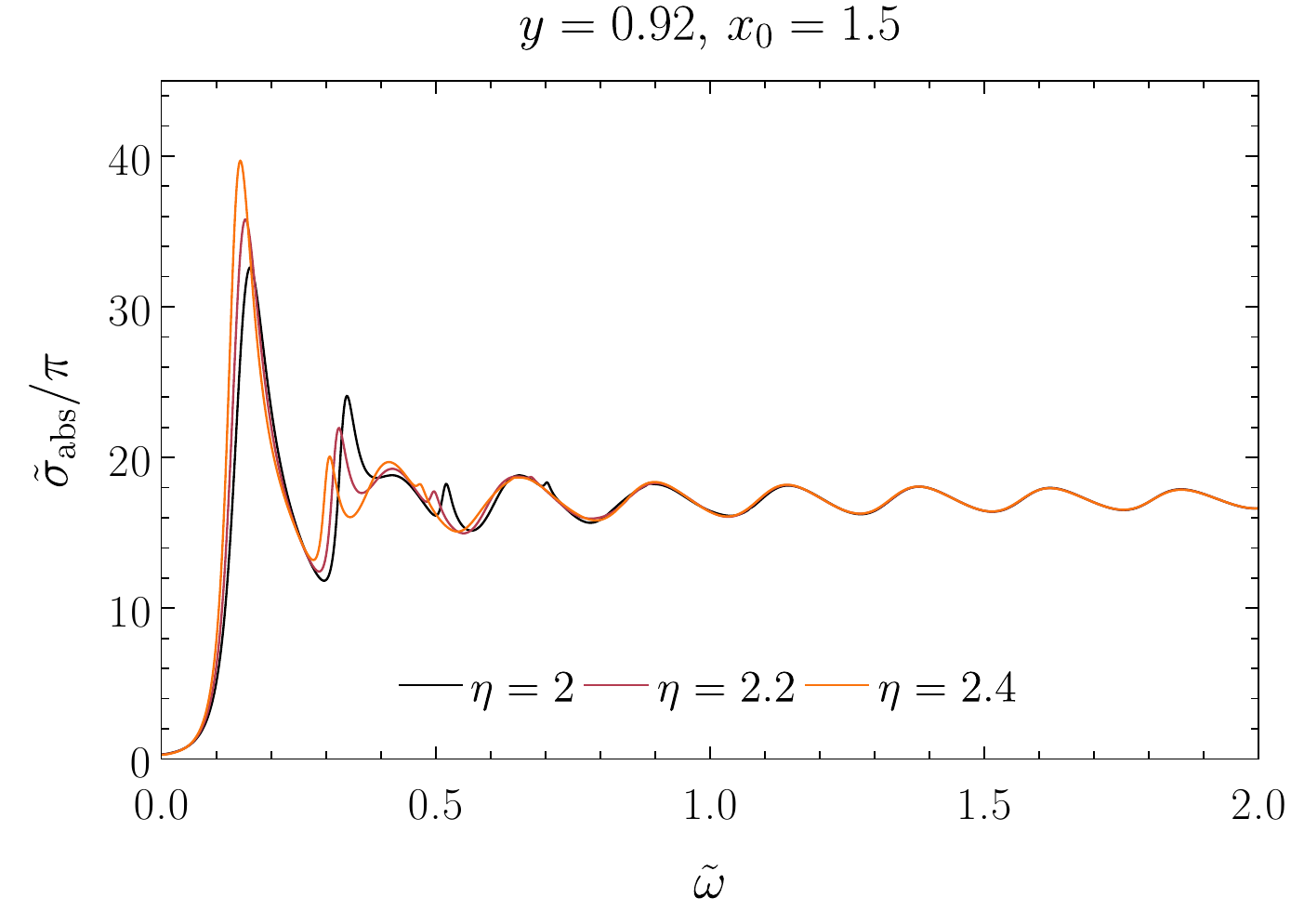}\hspace{0.3cm} \includegraphics[width=\columnwidth]{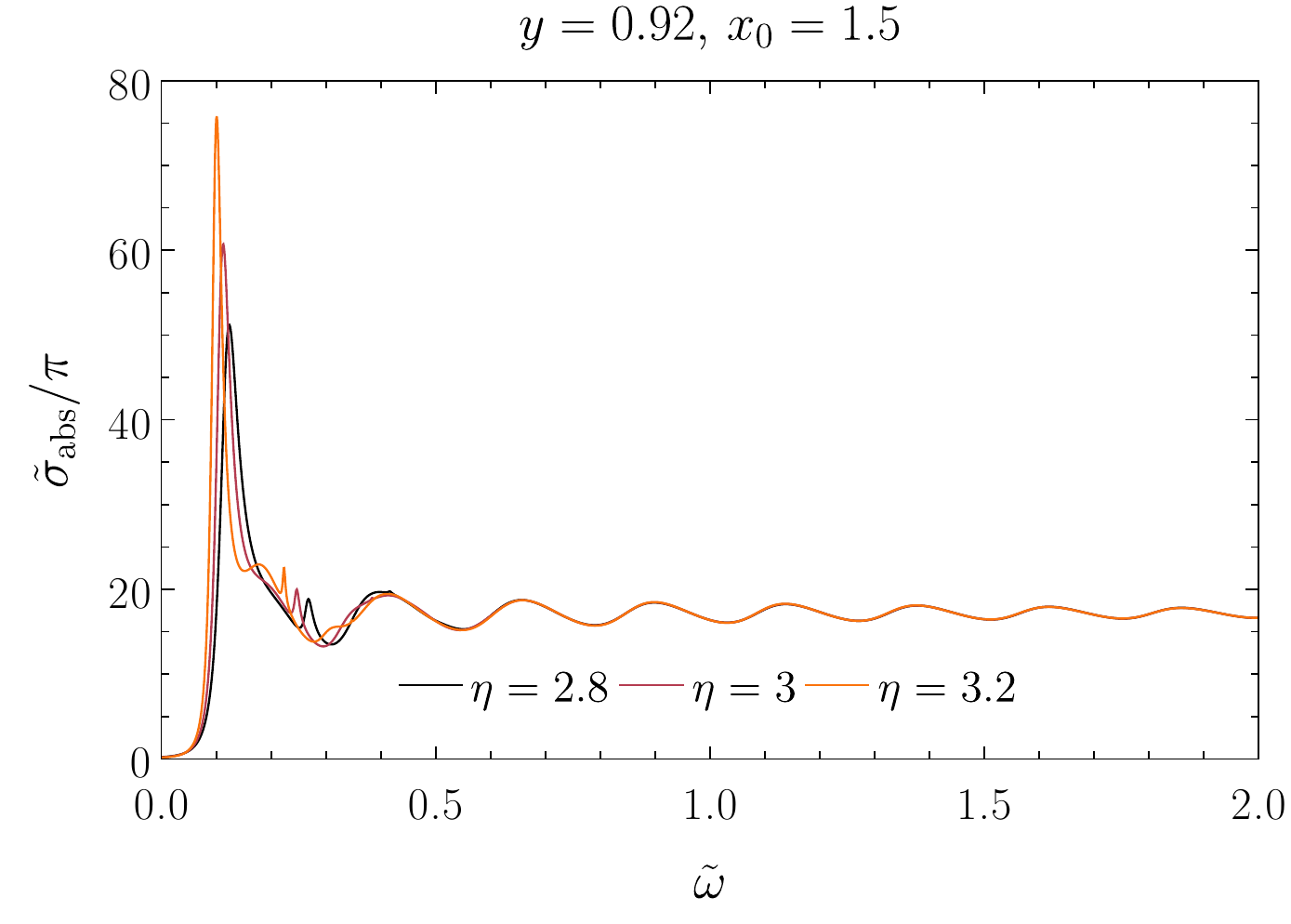}
		\caption{Absorption spectra of RN-AWHs with $x_0=1.5$ and $y=0.92$, supported by SPE (top panel) and SNE (bottom panel) thin shells.}
		\label{fig:absorption_spectra_RNAWH2}
	\end{figure}

	Since $0<y<1$, the absorption spectra of RN-AHWs goes to the RN one in the eikonal limit, and the effects of the other parameters, namely $x_0$ and $\eta$, are less relevant in the limit that $\tilde{\omega}\gg 1$. However, for low-energy waves the absorption profile has a deep dependence on the shell location, $x_0$, and on the charge-to-charge ratio between the wormhole sides, $\eta$. In order to investigate how the parameters beyond the charge-to-mass ratio influence the absorption process, let us first fix the dimensionless charge and the shell location, namely let us set $y=0.92$ and $x_0=1.5$. For fixed values of $x_0$ and $y$, the energy parameter of the shell decreases as we increase the charge-to-charge ratio $\eta$. Then, one finds that with this choice of parameters, wormholes are supported by SPE shells if $1.98959<\eta<2.78261$ and are supported by SNE shells if $2.78261<\eta<3.27128$. 
	In Fig.~\ref{fig:absorption_spectra_RNAWH2} we plot SPE (top panel) and SNE (bottom panel) configurations with $x_0=1.5$ and $y=0.92$ for different choices of $\eta$. In the high-frequency regime the total absorption cross section oscillates around the classical absorption cross section, as in the RN case, and the role of $\eta$ is less relevant. However, we notice that the charge-to-charge ratio strongly affects the absorption spectra for $\tilde{\omega}<1$. 
	From Fig.~\ref{fig:absorption_spectra_RNAWH2} we notice that the narrower peaks that arise in the wormhole absorption spectra are shifted to the left as we increase the value of $\eta$, and the first peak gets higher for greater values of $\eta$, regardless of the sign of the shell's energy-density. The behavior of the other narrower peaks is different depending on the energy content of the shell, namely: (i) if the shell has a positive-energy density, then increasing the charge-to-charge ratio (which corresponds to decrease the energy parameter $\tilde{\gamma}$) diminishes the other narrower peaks; (ii) if the shell has negative-energy density, increasing $\eta$ (consequently going to more negative values of $\tilde{\gamma}$) also increases the other narrower peaks. 
	
\begin{figure}
		\centering
		\includegraphics[width=\columnwidth]{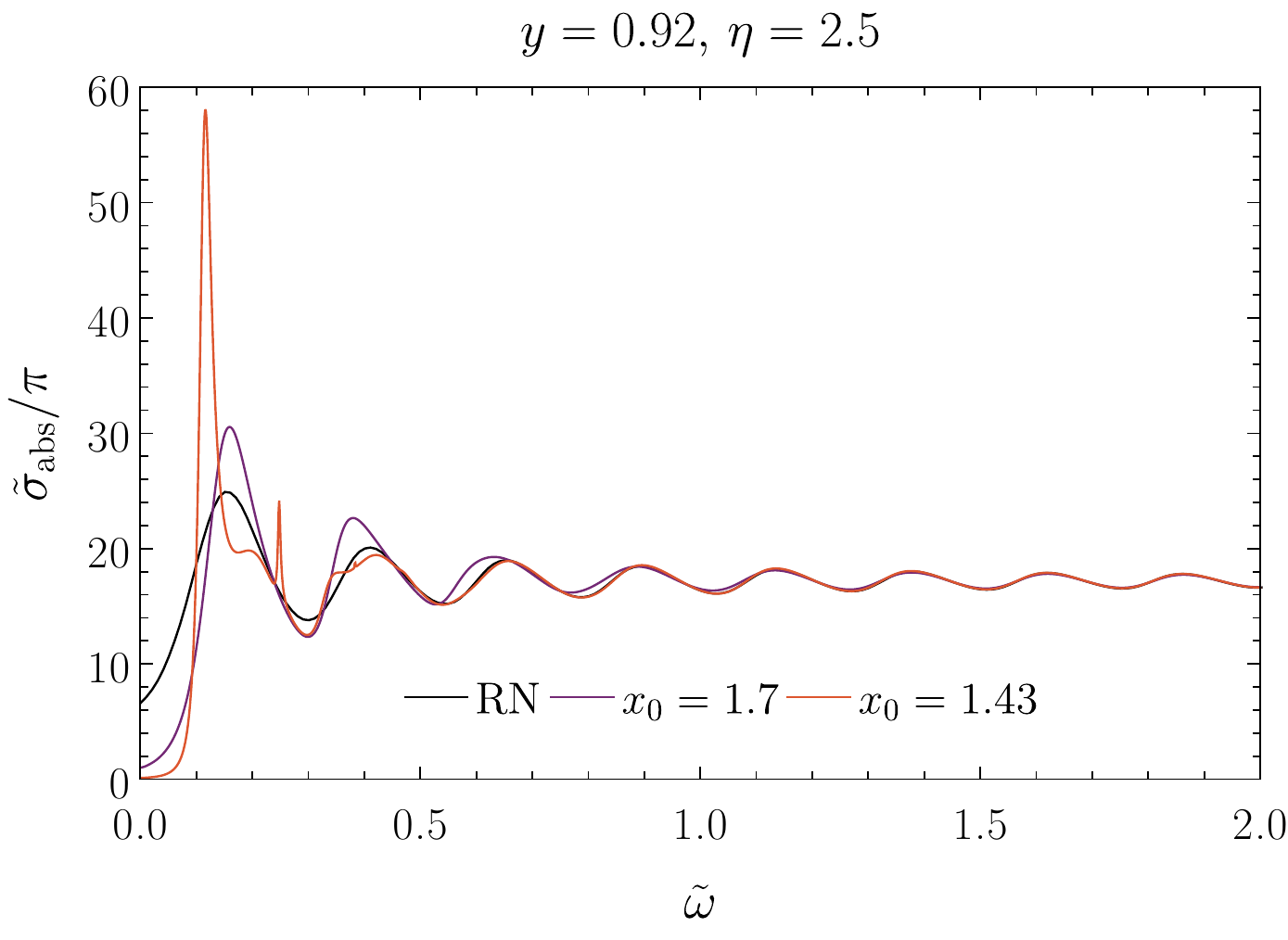}\hspace{0.3cm}
		\caption{Absorption spectra of RN-AWHs with $y=0.92$ and $\eta=2.5$ and some choices of $x_0$ compared with the absorption of a RN black hole with the same dimensionless charge as ${\cal M}_-$.}
		\label{fig:absorption_spectra_RNAWH3}
	\end{figure}

	In order to investigate how the shell location affects the absorption cross section, let us first consider a configuration with dimensionless charge $y<1$. In Fig.~\ref{fig:absorption_spectra_RNAWH3} we plot the total absorption cross section for some RN-AWHs with $y=0.92$, $\eta=2.5$ and different shell locations, $x_0$. We also compare the absorption spectra of those AWH configurations with the one of a RN black hole with the same value as dimensionless charge. Again, we notice that in the high-frequency regime, the total absorption cross section of these configurations presents the RN profile, oscillating around the classical absorption cross section. However, in the low-frequency regime, the behavior of the absorption spectra is significantly modified by the shell location, since the shape of the effective potential is particularly dependent on the throat location. From Fig.~\ref{fig:absorption_spectra_RNAWH3}, we notice that configurations with smaller shell radius have bigger absorption peaks, slightly shifted to the left, in the low-frequency regime, and may exhibit additional sharp peaks as $x_0$ diminishes.
	
\begin{figure}
		\centering
		\includegraphics[width=\columnwidth]{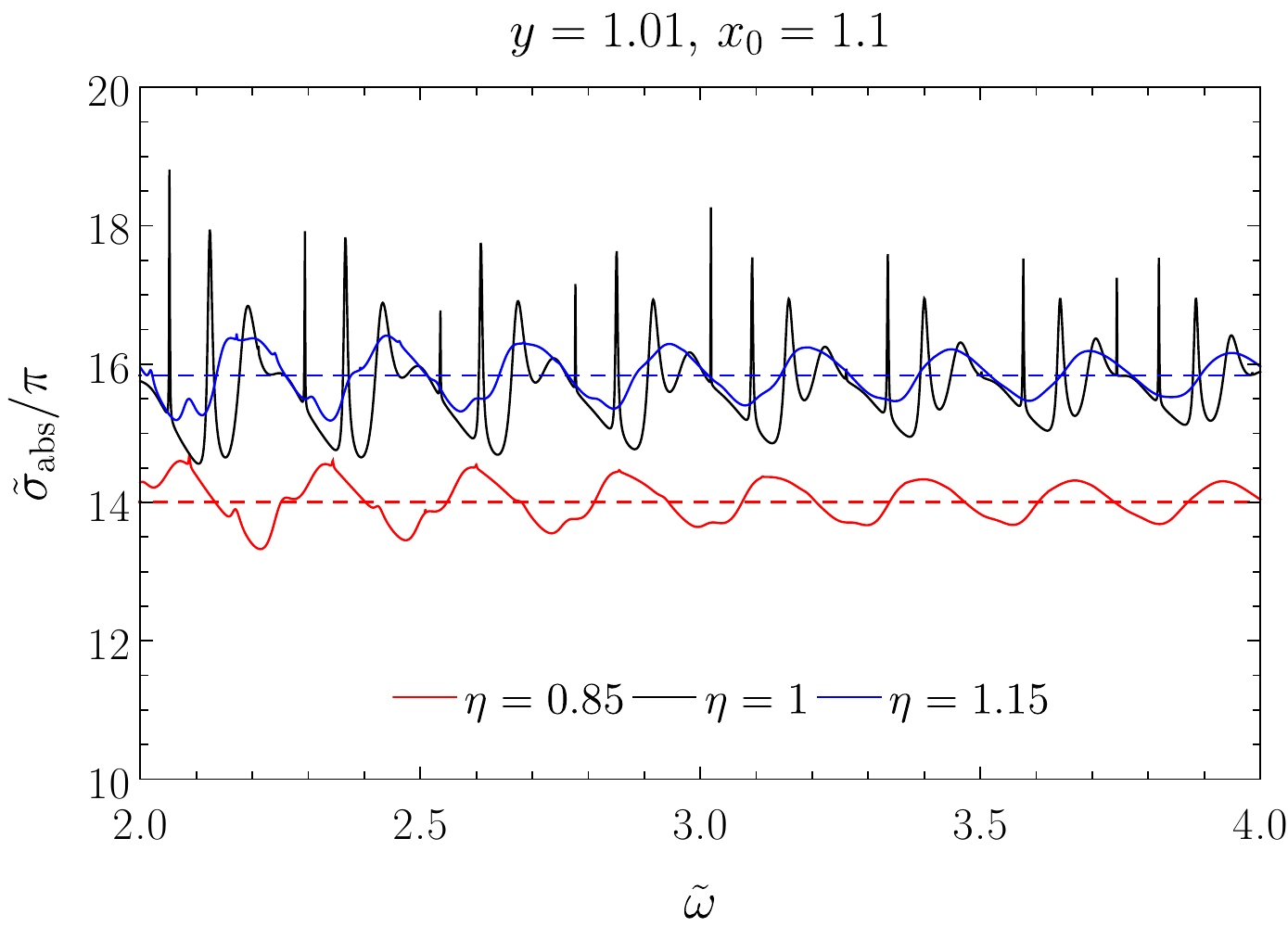}\hspace{0.3cm}
		\caption{High-frequency absorption spectra of wormholes with $x_0=1.1$ and $y=1.01$ and some choices of $\eta$. The dashed lines represent the classical absorption cross sections for each spacetime.}
		\label{fig:absorption_spectra_hf}
	\end{figure}

	When considering $y>1$, both $\eta$ and $x_0$ may remarkably affect the absorption spectra. If ${\cal M}_-$ is a naked RN spacetime, then the charge-to-charge ratio may be $\eta<1$, which leads to an effective potential with a higher peak in ${\cal M}_{+}$ instead of ${\cal M}_-$, and consequently with a total absorption cross section, in the high-frequency regime, smaller than for $\eta\geq 1$. We show this behavior in Fig.~\ref{fig:absorption_spectra_hf} where we plot the high-frequency regime of the total absorption cross section of a symmetric wormhole $(\eta=1)$ and two asymmetric wormholes ($\eta=0.85$ and $\eta=1.15$) with $x_0=1.1$ and $y=1.01$. It is important to point out that in both asymmetric configurations the presence of sharp peaks is attenuated in the high-frequency regime when compared with a symmetric configuration. The role of the symmetry in the presence of the sharp peaks will be discussed in the next section.
	
\begin{figure}
		\centering
		\includegraphics[width=\columnwidth]{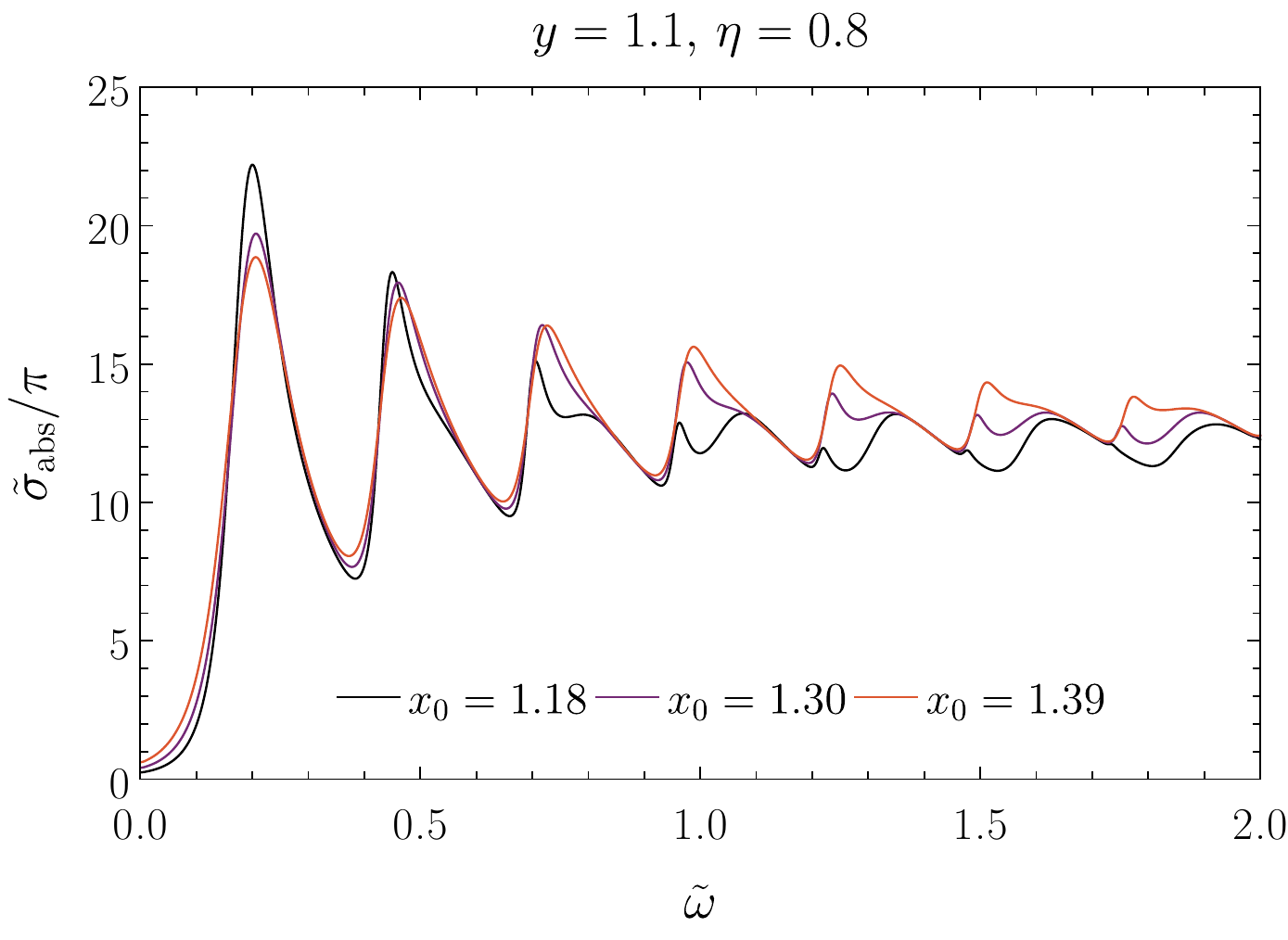}\hspace{0.3cm}
		\caption{Absorption spectra of RN-AWHs with $y=1.1$ and $\eta=0.8$ and some choices of $x_0$.}
		\label{fig:absorption_spectra_RNAWH4}
	\end{figure}

	The throat location, $x_0$, may also imply interesting features when $y>1$. In Fig.~\ref{fig:absorption_spectra_RNAWH4} we plot the total absorption cross section for some RN-AWH with $y=1.1$, $\eta=0.8$ and different values of the shell radius, $x_0$. We notice that, differently from the $y<1$ case, the shell radius plays a non-negligible role in the moderate-to-high frequency regime. We see that smaller values of $x_0$ present bigger absorption peaks in the low-frequency regime. However, for moderate-to-high frequencies the absorption peaks of those configurations are smaller if compared with the ones of bigger shell radius. Additionally, wormholes with smaller shell radius may present new absorption peaks in the high frequency regime, differently from the $y<1$ case, where the new peaks appear usually in the low-frequency region.	

	\subsection{Quasibound states}\label{sec:qbs}
	A remarkable feature that appears in wormhole scenarios is the existence of sharp peaks in the absorption spectra. These peaks are associated with the shape of the effective potential. If the effective potential has a valley, quasibound states can exist around the wormhole throat, producing resonant amplifications in the absorption cross section. These modes are similar to the trapped modes found in ultracompact objects~\cite{macedo:2018}, which in the eikonal limit may be associated with a stable light ring~\cite{cardoso:2014b}. 
	
	The quasibound states are characterized by complex frequencies with small imaginary part. In order to find the trapped modes one considers the boundary conditions
	\begin{equation}
		\psi(\xst)\sim\left\{
		\begin{array}{ll}
			e^{i \tilde{\omega} \xst},&\xst\to+\infty, \\
			e^{-i\tilde{\omega} \xst},& \xst\to-\infty.
		\end{array}\right.\label{eq:trapped_modes}
	\end{equation}
	These boundary conditions generate an eigenvalue problem to $\tilde{\omega}$, and one may apply standard methods to determine those frequencies. From an approximation based on the Breit-Wigner expression for nuclear scattering~\cite{breit:1936,feshbach:1947}, one can relate the grey-body factor with the trapped modes, namely~\cite{macedo:2018}
	\begin{equation}
		\label{eq:Breit-Wigner}
		|{\cal T}_{\tilde{\omega} \ell}|^2\Big\vert_{\tilde{\omega}\approx\tilde{\omega}_r} \propto \dfrac{1}{(\tilde{\omega}-\tilde{\omega}_r)^2	+\tilde{\omega}_i^2}.
	\end{equation}
	Hence, one notices that the position of the resonant peaks in the transmission coefficients is determined by the real part of the mode, $\tilde{\omega}_r$, while the imaginary part, $\tilde{\omega}_i$, determines the sharp shape and height of the peaks.
	
\begin{figure*}
		\centering
		\hspace{-1cm}\includegraphics[width=\columnwidth]{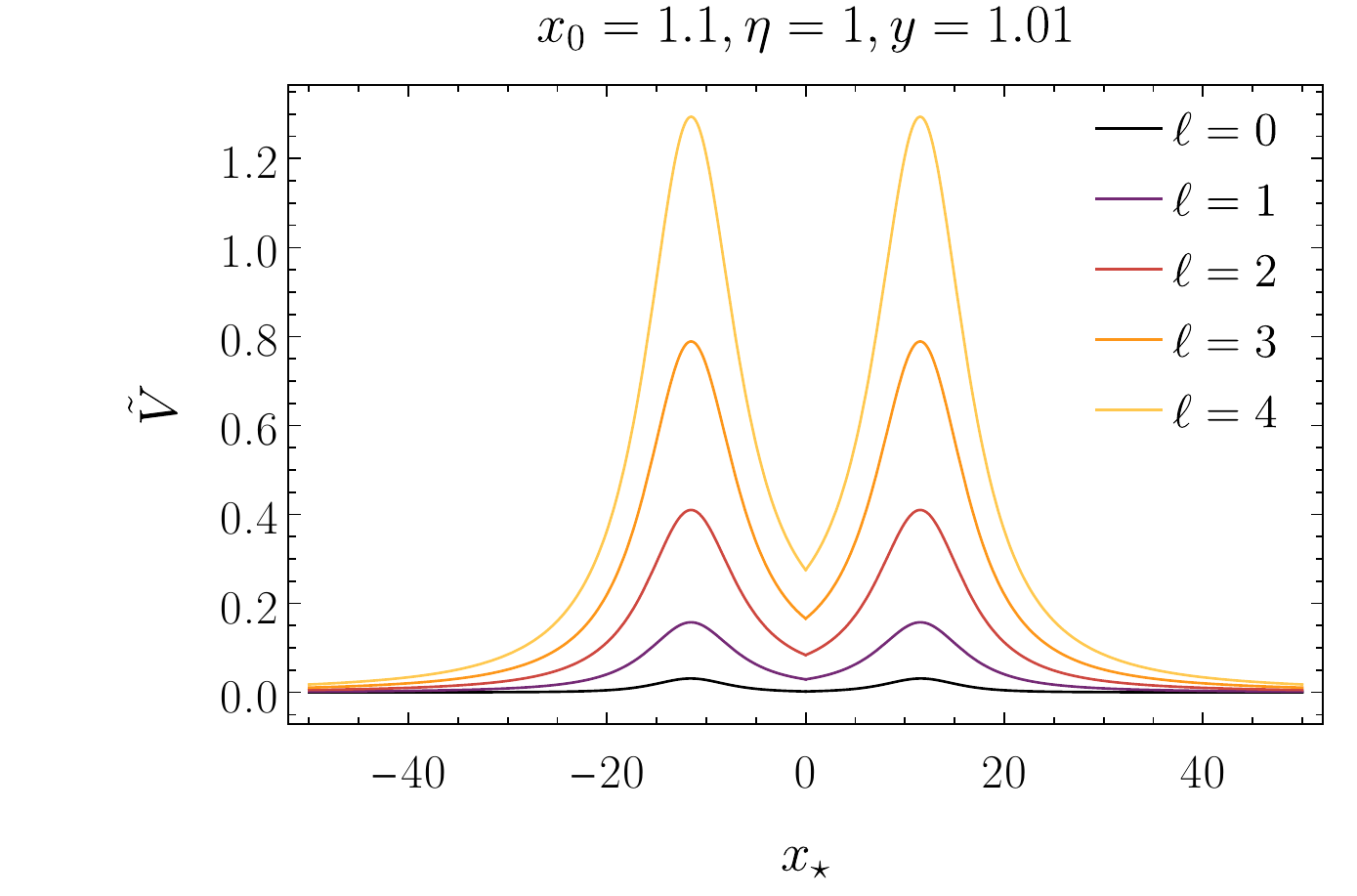}
		\includegraphics[width=\columnwidth]{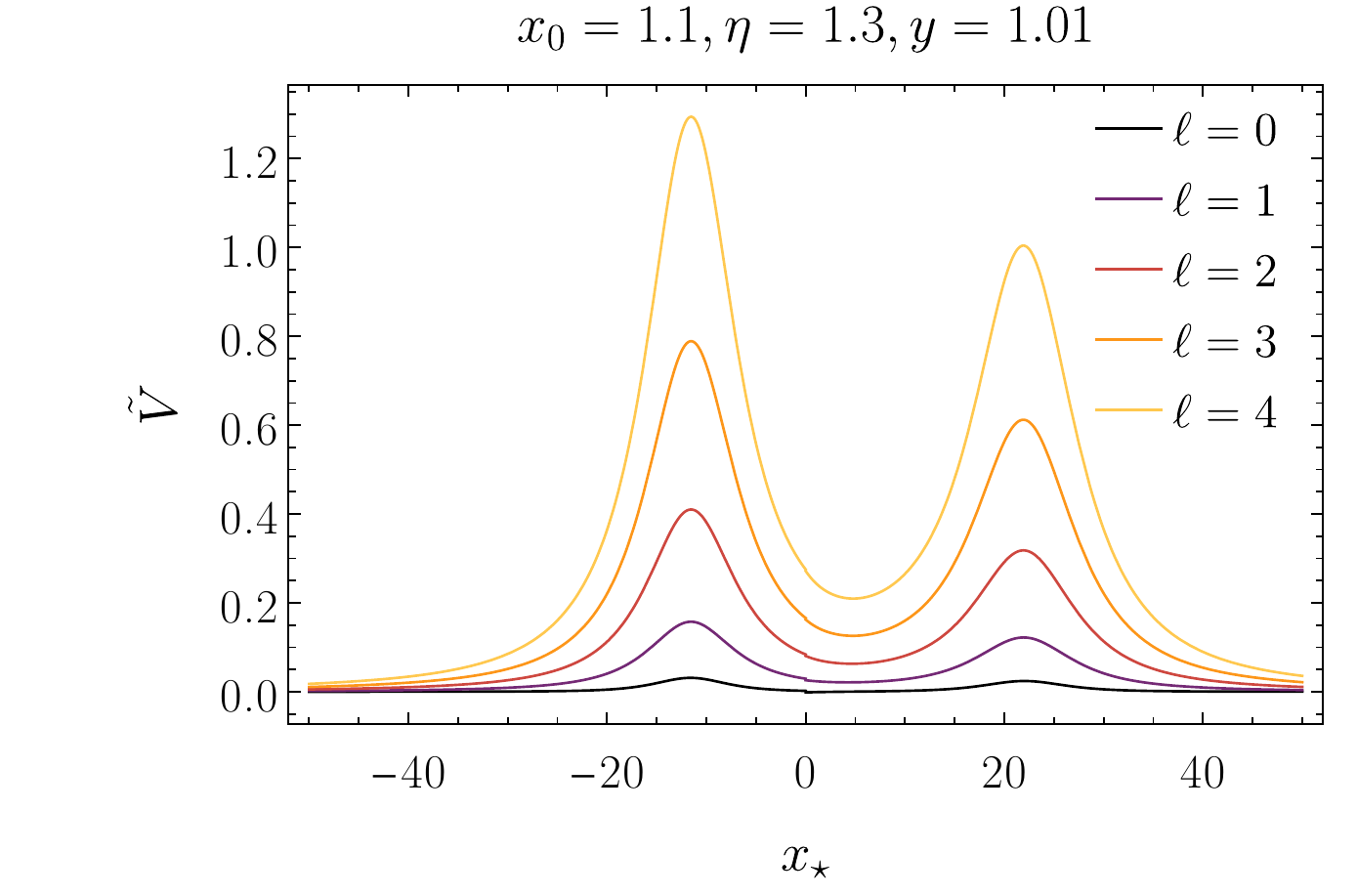}

		\hspace{-1cm}\includegraphics[width=\columnwidth]{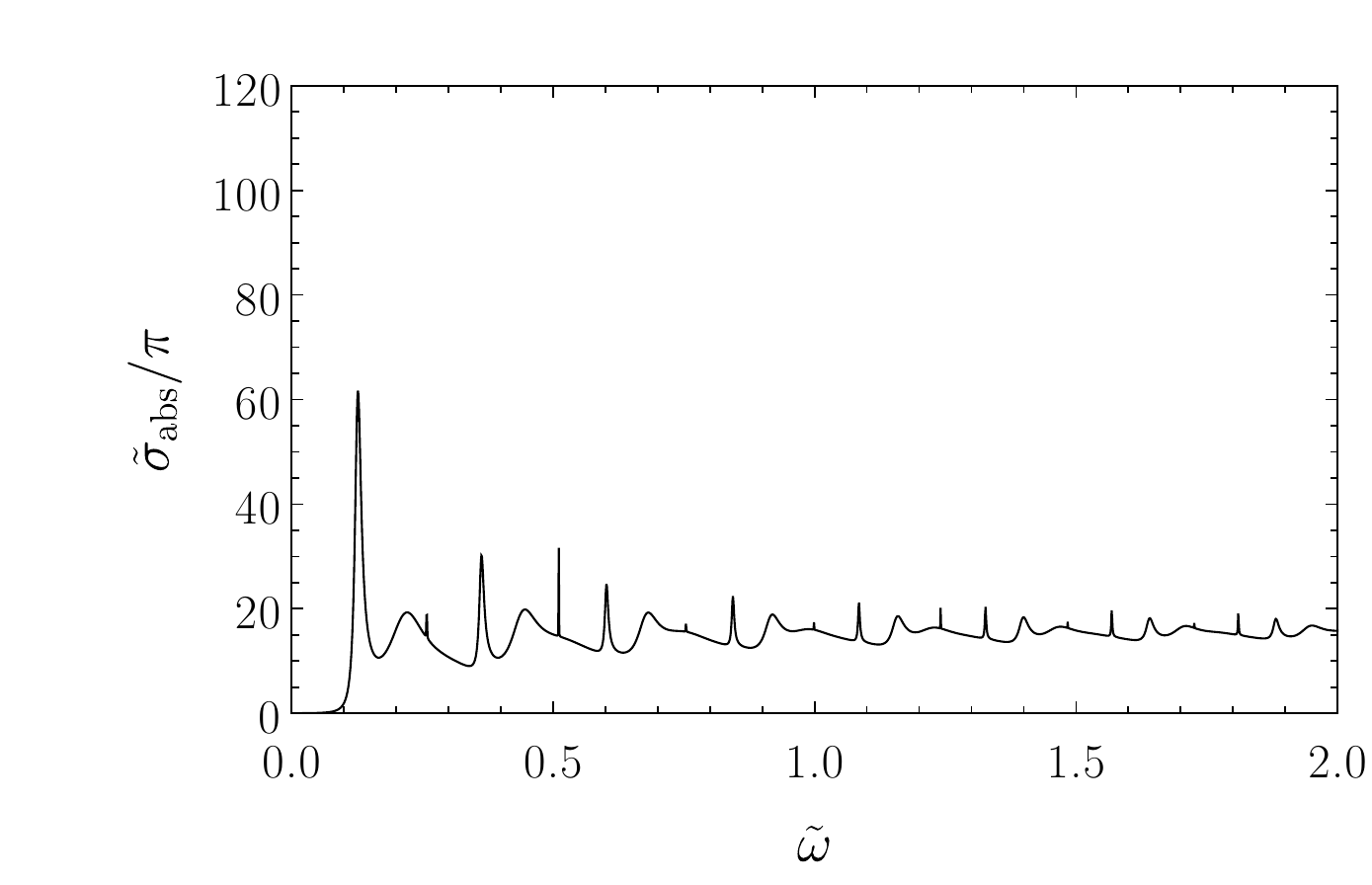}
		\includegraphics[width=\columnwidth]{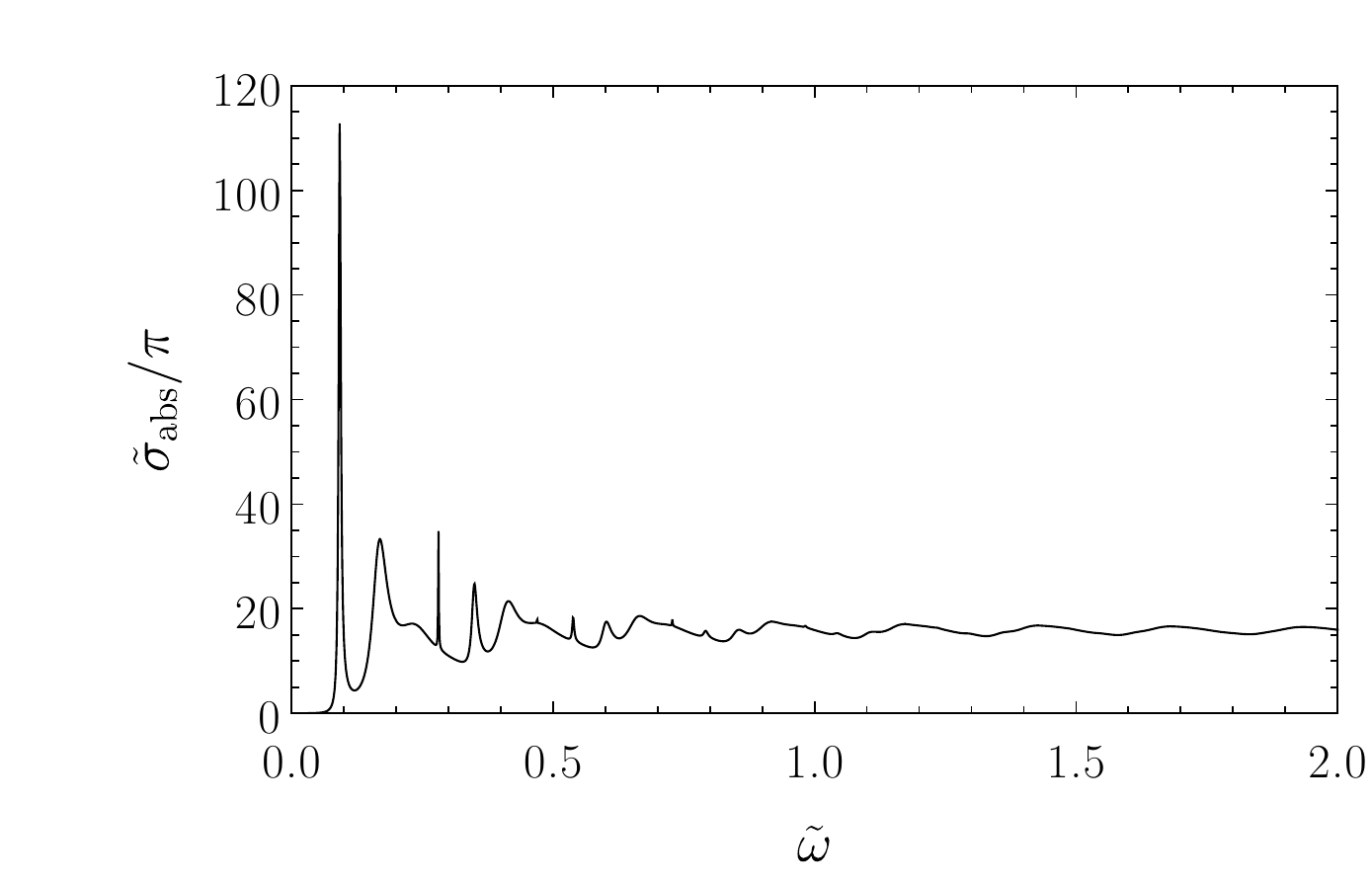}

		\hspace{-1.1cm}\includegraphics[width=\columnwidth]{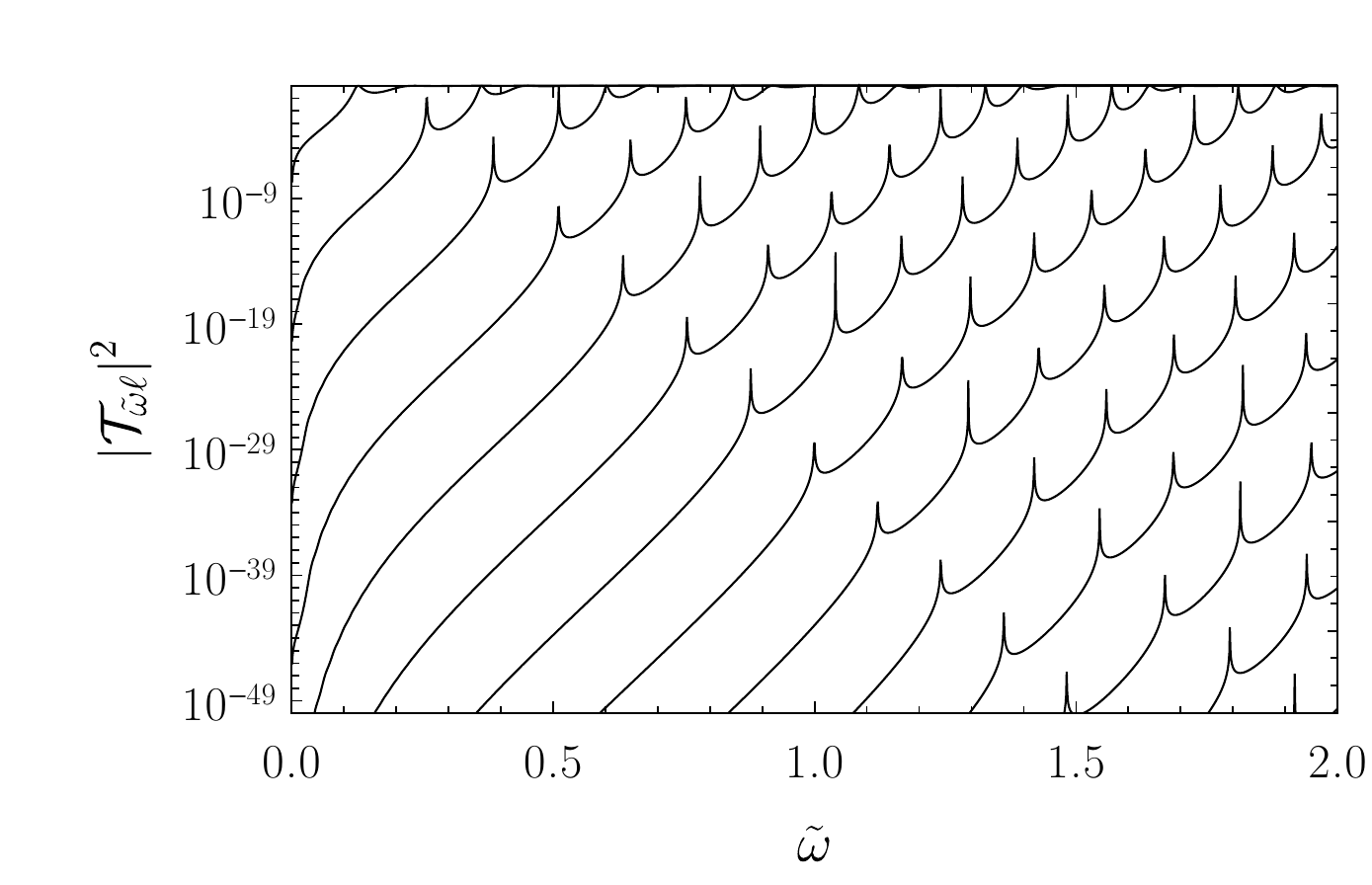}
		\includegraphics[width=\columnwidth]{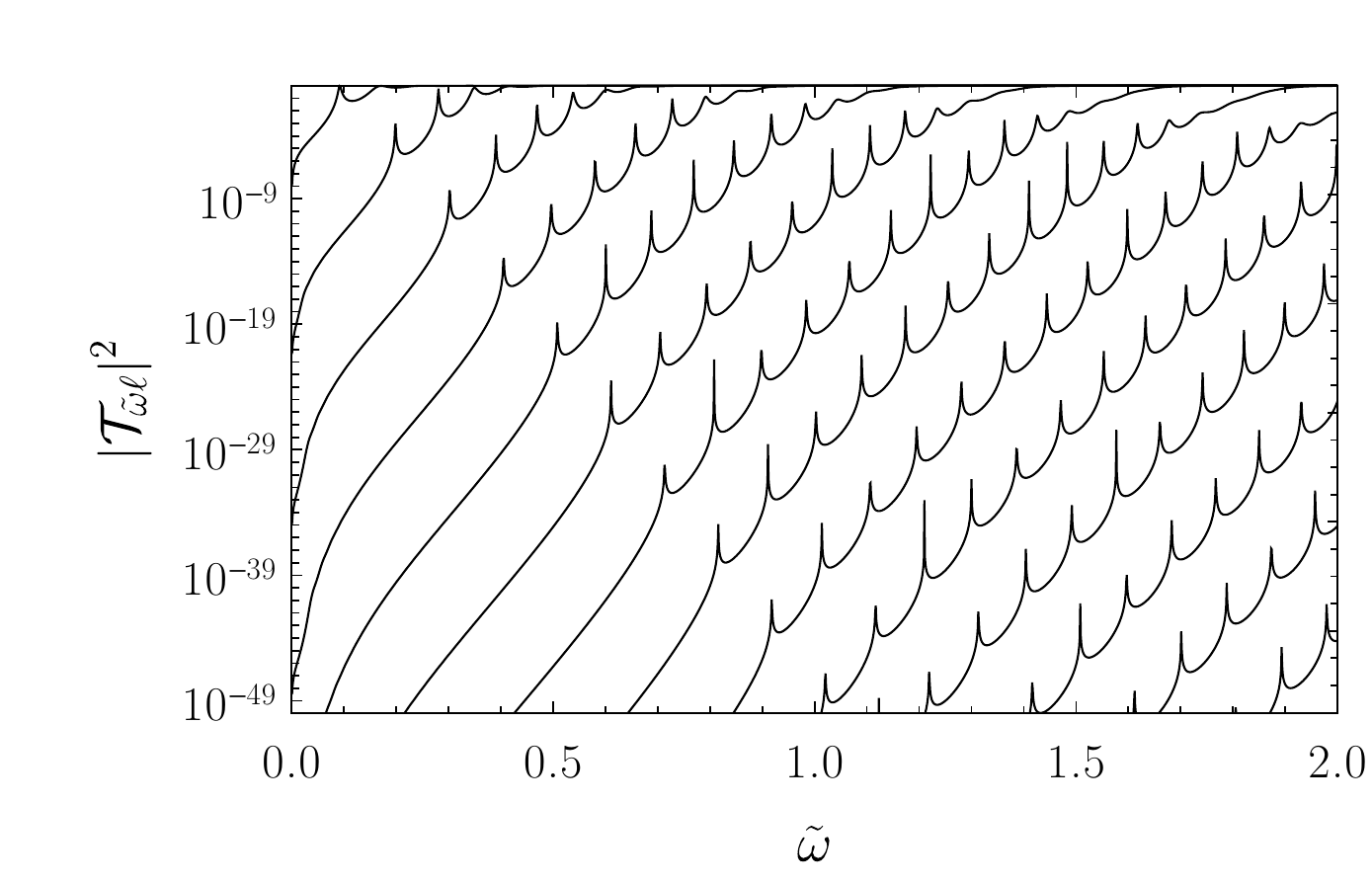}
		\caption{Effective potential (top row), absorption spectra (middle row) and transmission coefficients (bottom row) of a RN symmetric wormhole (left column) and of a RN-AWH with $\eta=1.3$ (right column). Both wormholes have $y=1.01$ and $x_0=1.1$.}
		\label{fig:abs_tra_RNSWH}
	\end{figure*}

Due to the freedom that we have to construct wormholes in Palatini $f({\cal R})$ gravity, the effective potential of those configurations may present different \textit{asymmetries} (cf. Fig.~\ref{fig:effective_potential_RNAWH}), which, as we saw in the previous section, may lead to more or less additional peaks in the absorption spectra. In Fig.~\ref{fig:abs_tra_RNSWH} we plot the effective potential (top row), the total absorption cross section (middle row) and the transmission coefficients (bottom row) of a symmetric (left column) and an asymmetric (right column) wormhole supported by SNE shells. We notice that both configurations present sharp peaks for $\tilde{\omega}<1$. By using the direct integration method, and a standard root-finder method, one can find the frequencies that solve the eigenvalue problem and characterize the trapped modes. In Table ~\ref{table1} we present some trapped modes for the asymmetric wormhole considered in Fig.~\ref{fig:abs_tra_RNSWH}.
	\begin{table}[hbtp!]
		\centering \caption{Trapped modes frequencies for RN-AWHs.}
		\vskip 10pt
		\begin{tabular}{@{}ccccccc@{}}
			\hline \hline
			\multicolumn{3}{c}{$x_0=1.1$, $y=1.01$, $\eta=1.3$}\\ \hline\hline
			$\ell$\hspace{1cm}      &$\tilde{\omega}_r$\hspace{1cm} &$\tilde{\omega}_i$ \\
$0$\hspace{1cm}            &$ 0.0919 $\hspace{1cm}         		& $ -2.4781\times 10^{-3}$\\
\hspace{1cm}               &$ 0.1707  $\hspace{1cm}          &$ -8.7901\times 10^{-3} $        \\
\hspace{1cm}               &$ 0.2535  $\hspace{1cm}          &$ -9.2636\times 10^{-3} $        \\
\hspace{1cm}               &$ 0.3404  $\hspace{1cm}          &$ -9.3142\times 10^{-3} $        \\
\hspace{1cm}               &$ 0.4401  $\hspace{1cm}          &$ -9.8017\times 10^{-3} $        \\
\hspace{1cm}               &$ 0.5155  $\hspace{1cm}          &$ -9.9624\times 10^{-3} $        \\
			\hline
			$1$\hspace{1cm}               &$ 0.1985  $\hspace{1cm}          &$ -1.5661\times 10^{-5} $        \\
			\hspace{1cm}               &$ 0.2809  $\hspace{1cm}          &$ -6.4116\times 10^{-4} $        \\
			\hspace{1cm}               &$ 0.3500  $\hspace{1cm}          &$ -6.1025\times 10^{-3} $        \\
			\hspace{1cm}               &$ 0.4176  $\hspace{1cm}          &$ -6.3230\times 10^{-3} $        \\
			\hspace{1cm}               &$ 0.5753  $\hspace{1cm}          &$ -6.6483\times 10^{-3} $        \\
						\hspace{1cm}               &$ 0.6614  $\hspace{1cm}          &$ -6.7927\times 10^{-3} $        \\
			\hline
			$2$\hspace{1cm}               &$ 0.3024  $\hspace{1cm}          &$ -4.6635\times 10^{-8} $        \\
			\hspace{1cm}               &$ 0.3907  $\hspace{1cm}          &$ -4.6659\times 10^{-6} $        \\
			\hspace{1cm}               &$ 0.4695  $\hspace{1cm}          &$ -1.5562\times 10^{-4} $        \\
			\hspace{1cm}               &$ 0.5383  $\hspace{1cm}          &$ -2.1412\times 10^{-3} $        \\
			\hspace{1cm}               &$ 0.6023  $\hspace{1cm}          &$ -1.7040\times 10^{-2} $        \\
			\hspace{1cm}               &$ 0.6722  $\hspace{1cm}          &$ -1.7339\times 10^{-2} $        \\
			\hline \hline
		\end{tabular}
		\label{table1}
	\end{table}  
	
For the asymmetric configuration, as the frequency increases we can barely see the additional peaks in the absorption spectra. This can be understood by analyzing the asymmetry in the effective potential as one increases the $\ell$ number. For lower values of the angular momentum mode, the difference between the heights of the peaks in the effective potential is much smaller when compared with that difference for large values of $\ell$. Hence, low-frequency trapped modes lie between two potential peaks with almost the same height, creating the resonances in the absorption spectra for $\tilde{\omega}<1$. For larger values of $\ell$ the difference between the heights of the effective potential peaks becomes non-negligible, and since in the high-frequency regime the higher effective potential peak determines the absorption behavior, we almost do not see resonant amplifications in the absorption cross section. However, a glance at the transmission coefficients of the asymmetric configuration shows that we still have trapped modes around the wormhole in the eikonal limit, notwithstanding we can not see the resonant peaks in the absorption profile. Therefore, the presence of spectral lines is more evident as one restores the symmetry of the thin shell wormholes.
		
\section{Discussion and Final remarks}\label{sec:con}
We have studied the absorption properties of stable RN-AWHs constructed via the thin-shell formalism in Palatini $f(\cal R)$ gravity. The configurations considered in this paper are either a RN black hole spacetime glued to a RN naked singularity spacetime or two RN naked singularities spacetimes glued together. The latter represents a novel configuration within the wormhole literature.

The Palatini formulation and the (usual) metric formulation of General Relativity yield the same field equations and junction conditions. However, beyond Einstein's gravity (for nonlinear Lagrangians) this equivalence no longer holds, resulting in different field equations and junction conditions. The distinct junction conditions imply that all discussions regarding the stability of thin shell solutions must be reconsidered within the $f(\cal R)$ framework. In fact, this led to the discovery of stable solutions with positive energy density at the matching surface, something not allowed according to the standard General Relativity junction conditions.

By analyzing the massless scalar field in the vicinity of stable RN-AWHs (see Fig.~\ref{fig:space_of_param}), we found basically four effective potential behaviors, namely: (i) two smooth peaks connected by a discontinuous valley (for all values of $\ell$); (ii) a single smooth peak in the effective potential for $\ell\geq 1$, and a discontinuous valley for $\ell=0$; (iii) a discontinuous sharp peak at the throat for $\ell\geq 1$, and a discontinuous well for $\ell=0$; and (iv) a sharp discontinuous peak at the throat followed by a smooth valley and a smooth peak on each side for $\ell\geq 1$ (for $\ell=0$ the sharp discontinuous peak is replaced by a discontinuous well). Cases (i) and (ii) occur when one BH spacetime is used to build the wormhole, while cases (iii) and (iv) occur when both sides are composed by naked singularity spacetimes. Since the shape of the effective potential varies considerably with the chosen parameters, one expects noticeable changes in the absorption profile of RN-AWHs. In order to investigate how the throat location and the charge values influence the absorption, we analyzed several SPE and SNE configurations.

If the effective potential has a valley (continuous or discontinuous), quasibound states emerge around the throat of the wormhole. These quasitrapped modes create resonances in the absorption spectra (sharp peaks appear in the absorption cross section), which make the RN-AWH absorption profiles very different from the ones of RN black holes. These new peaks are highly influenced by the symmetry of the potential well. If the effective potential exhibits a symmetric valley, the resonances are noticeable in the whole range of frequency. On the other hand, the presence of an asymmetry in the effective potential results in the attenuation of the resonant peaks associated with quasibound states for higher $\ell$-modes. Thus, the high-frequency regime of the absorption cross section becomes degenerate with the prediction for the standard RN black hole, in contrast to symmetric wormhole configurations. Consequently, even minor deviations from symmetry in wormhole spacetimes can yield significant differences on the observable characteristics associated to quasibound states. Therefore, the RN-AWH can present a remarkable absorption profile compared with previous results where those spectral lines were found (see, for instance, Ref.~\cite{macedo:2018} for resonant peaks in ultracompact objects and Refs.~\cite{delhom:2019,limajr:2020} for resonant peaks in solutions that interpolate between black holes and compact objects with wormhole topology). If the resonances of the symmetric case persist at very high frequencies, one could expect nontrivial effects even in the geometrical optics approximation. This could lead to unexpected features in gravitational waves spectra and electromagnetic shadows. Further research in this direction is currently underway.

By considering an asymmetric configuration with  dimensionless charge $y<1$, the wormhole can mimic the standard RN black hole absorption. This can be understood by the fact that the total absorption cross section depends on the dominant light ring (associated with the highest peak of the effective potential in the eikonal limit~\cite{cilasjr:2021}). Therefore, by grafting a RN black hole (before its photon sphere) with a RN naked singularity spacetime, the dominant light ring will be the one of the RN black hole and it will dictate the absorption cross section profile in the eikonal limit. By restoring the symmetry ($\eta\to 1$), although the absorption cross section oscillates around the classical value of RN black hole, it can be distinguished by the presence of the spectral lines. 

An interesting and, to our knowledge, new absorption behavior appears when we cut and paste two naked singularities, presenting discontinuous sharp peaks at the throat in the effective potential. In these scenarios, the throat acts like an effective photon sphere~\cite{wang:2020,shaikh:2019}. Since, at the throat, the effective potential reaches a maximum value, the effective light ring related to it is the dominant light ring; therefore, the total absorption cross section will go to the area of the shadow associated with the effective light ring. Interestingly, the oscillatory pattern of the absorption profile rapidly attenuates and the absorption cross section slowly goes to the shadow area associated with null geodesics trapped on the throat.

The SPE configurations studied here present a low-frequency limit of the total absorption cross section much smaller than the corresponding black hole ones, analogously to other wormhole cases previously analyzed~\cite{delhom:2019,limajr:2020}. 
On the other hand, although several SNE configurations also present this typical almost zero low-frequency regime, we found configurations supported by negative energy shells such that, as we intensify the charge contents on both sides of the wormhole, the total absorption cross section noticeably increases in the low-frequency regime. 
This indicates that the low-frequency absorption properties are sensitive to the wormhole model parameters, having no obvious trend, which might be related with the presence of a discontinuous well in the effective potential for $\ell=0$.  
A deeper analysis concerning the zero-frequency limit of the total absorption cross section of wormholes should be done in order to better understand those features.
	
Our results indicate that asymmetric wormholes may carry different observational imprints, compared to either black holes or symmetric wormholes. These asymmetries in the spacetime may lead, for example, to significant features in the quasi-normal mode spectrum and the possible presence of echoes. We are currently performing such investigations.
\begin{acknowledgments}
		The authors would like to acknowledge Funda\c{c}\~ao Amaz\^onia de Amparo a Estudos e Pesquisas (FAPESPA), Conselho Nacional de Desenvolvimento Cient\'ifico e Tecnol\'ogico (CNPq) and Coordena\c{c}\~ao de Aperfei\c{c}oamento de Pessoal de N\'ivel Superior (CAPES) -- Finance Code 001, from Brazil, for partial financial support.  
		This work has further been supported by the European Union’s Horizon 2020 research and innovation (RISE) programme H2020-MSCA-RISE-2017 Grant No. FunFiCO-777740 and by the European Horizon Europe staff exchange (SE) programme HORIZON-MSCA-2021-SE-01 Grant No. NewFunFiCO-101086251.
		AM-F and LC thank Universidade Federal do Par\'a, in Brasil and Universidade de Aveiro, in Portugal, respectively, for the kind hospitality.
		AM-F is supported by the Spanish Ministerio de Ciencia e Innovación with the PhD fellowship PRE2018-083802. This work is also supported by the Spanish Grant PID2020-116567GB- C21 funded by MCIN/AEI/10.13039/501100011033 and the project PROMETEO/2020/079 (Generalitat Valenciana). 
	\end{acknowledgments}


\appendix
\onecolumngrid
\bigskip	
	\section{Orthographic projection of the parameter space}
	\label{appendixOrtho}
	The orthographic projection is a common way to represent three dimensional objects in two dimensions. It is a representation of each side of an object as would be seen by an observer infinitely far away.
	
	\begin{figure*}[!h]
		\centering
		\includegraphics[width=\textwidth]{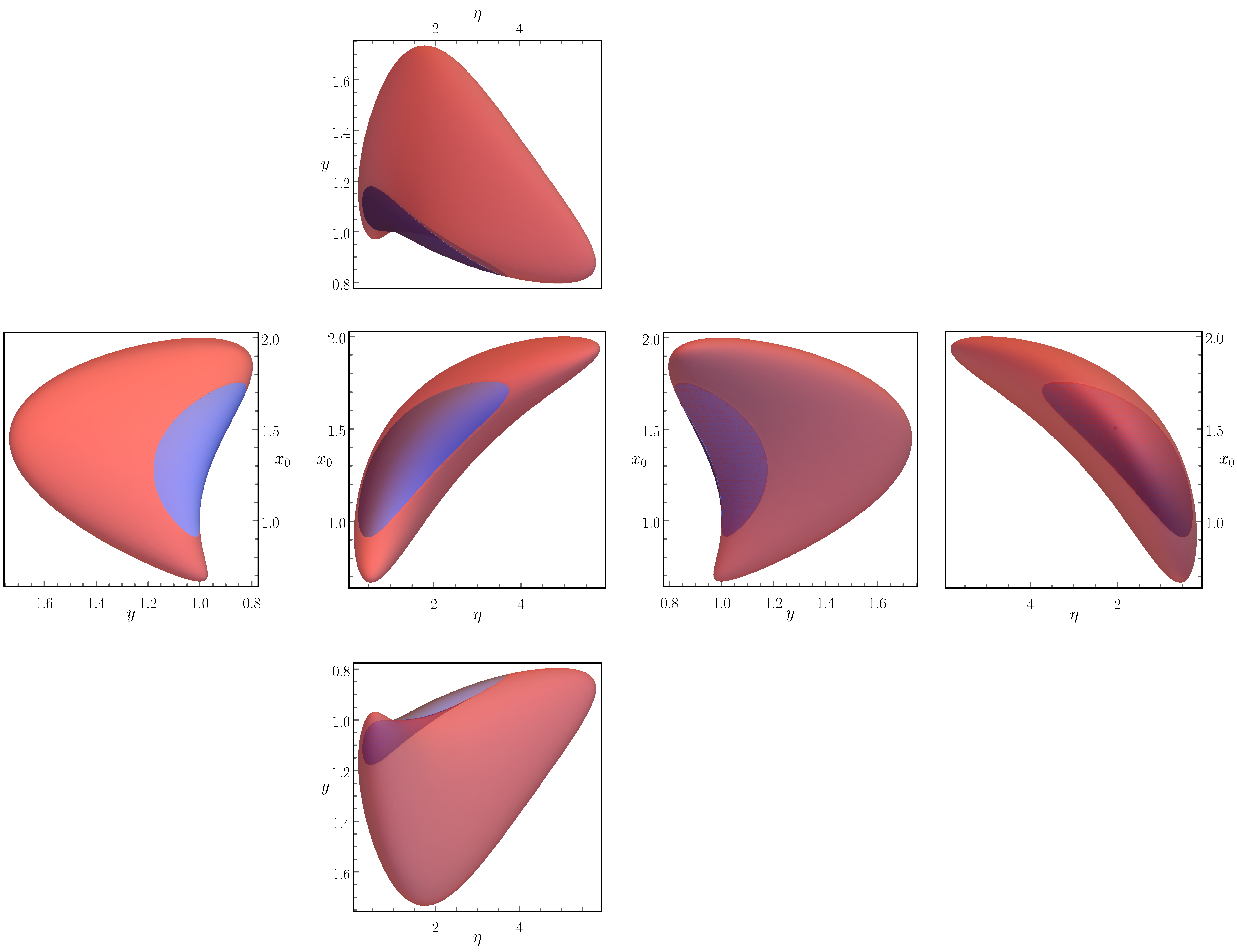}
		\caption{The six orthographic projections of the 3D plot shown in Fig.~\ref{fig:space_of_param}. The blue region represents the SPE space of parameters while the red region represents the SNE space of parameters. Transparency has been applied to the regions in order to make more noticeable that the SPE region (blue) is embedded onto the SNE region (red).}
		\label{fig:my_label}
	\end{figure*}
\twocolumngrid
		{}
\end{document}